\documentclass[12pt,superscriptaddress,amsmath,amssymb,nofootinbib,longbibliography,preprint,abstract]{revtex4-1}

\usepackage{graphicx}
\usepackage{dcolumn} 
\usepackage{bm}
\usepackage{amssymb}
\usepackage{amsmath}
\usepackage{epsfig}    
\usepackage{color}
\usepackage{slashed}
\usepackage[hypertex]{hyperref}
\begin{document} 

%\date{\today}
\preprint{OU-HET 1046}
\title{Radiative generation of neutrino masses in a 3-3-1 type model}

\author{Arindam Das}
\email{arindam.das@het.phys.sci.osaka-u.ac.jp}
\affiliation{Department of Physics, Osaka University, Toyonaka, Osaka 560-0043, Japan}

\author{Kazuki Enomoto}
\email{kenomoto@het.phys.sci.osaka-u.ac.jp}
\affiliation{Department of Physics, Osaka University, Toyonaka, Osaka 560-0043, Japan}

\author{Shinya Kanemura}
\email{kanemu@het.phys.sci.osaka-u.ac.jp}
\affiliation{Department of Physics, Osaka University, Toyonaka, Osaka 560-0043, Japan}

\author{Kei Yagyu}
\email{yagyu@het.phys.sci.osaka-u.ac.jp}
\affiliation{Department of Physics, Osaka University, Toyonaka, Osaka 560-0043, Japan}

\begin{abstract}

A new model for tiny neutrino masses is proposed in the gauge theory of $SU(3)_C \otimes SU(3)_L \otimes U(1)_X$, 
where neutrino masses are generated via the quantum effect of new particles.    
In this model, the fermion content is taken to be minimal to realize the gauge anomaly cancellation, 
while the scalar sector is extended from the minimal 3-3-1 model to have an additional $SU(3)_L$ triplet field. 
After $SU(3)_L\otimes U(1)_X$ is broken into $SU(2)_L\otimes U(1)_Y$, 
the ``Zee model'' like diagrams are naturally induced, which contain sufficient lepton flavor violating interactions to reproduce current neutrino oscillation data. 
Furthermore, the remnant $Z_2$ symmetry appears after the electroweak symmetry breaking, which guarantees the stability of dark matter. 
It is confirmed that this model can satisfy current dark matter data. 
As an important prediction to test this model, productions and decays of doubly-charged scalar bosons at collider experiments are discussed 
in successful benchmark scenarios. 

\end{abstract}
\maketitle
%\tableofcontents

\section{Introduction}

%[Overview]

The structure of the electroweak gauge symmetry $SU(2)_L \otimes U(1)_Y$ has been well established by various experiments. 
The spontaneous breaking $SU(2)_L \otimes U(1)_Y \to U(1)_{\rm em}$ by the Higgs mechanism has been confirmed 
by the discovery of the Higgs boson at the LHC. 
However, the possibility of larger gauge groups including the $SU(2)_L \otimes U(1)_Y$ symmetry can also be considered at higher energies than the 
electroweak scale. 
In fact based on various physics motivations, a plethora of models with extended gauge symmetries has been proposed.

One of the simple but well-motivated extensions is the $SU(3)_C \otimes SU(3)_L \otimes U(1)_X$ (3-3-1) gauge symmetry~\cite{Singer:1980sw,Valle:1983dk}. 
This extension can naturally give the answer to the question in the Standard Model (SM) ``Why are there three generations of chiral fermions?'' 
by the gauge anomaly cancellation. Namely, the number of generation has to be the same as the fundamental color degrees of freedom or its multiples~\cite{Frampton:1992wt}. 
There are many new particles in models with the 3-3-1 gauge symmetry. 
Clearly, these models necessarily introduce new gauge bosons. 
Cancellation of the gauge anomaly requires additional chiral fermions. 
In order to break the gauge symmetry into the SM one, additional scalar fields have also to be introduced. 
A next question is whether these new particles can play a role to explain phenomena which cannot be explained in the SM, such as 
the neutrino oscillation, the existence of dark matter and the baryon asymmetry of the Universe. 

Apart from the 3-3-1 models, a scenario generating tiny neutrino masses at quantum levels can naturally explain their smallness due to loop suppression factors without introducing super heavy new particles. 
The original model was proposed by Zee~\cite{Zee:1980ai}, in which neutrino masses are generated at 1-loop level. 
The model proposed by Zee and Babu~\cite{Zee:1985id,Babu:1988ki,Cheng:1980qt} generates neutrino masses at 2-loop level. 
After these models appeared, models with dark matter particles have also been proposed, 
in which the stability of dark matter is guaranteed by a discrete symmetry which is simultaneously forbid tree level diagrams for neutrino masses. 
For instance, the model proposed by Krauss-Nasri-Trodden~\cite{Krauss:2002px} and that by Ma~\cite{Ma:2006km} correspond to those along this line.
Therefore, if we can construct models for generating tiny neutrino masses radiatively based on the 3-3-1 scenario, 
we can explain origins of tiny neutrino masses, dark matter and generation of chiral fermions simultaneously. 
This is the subject of the present paper. 

Several models with radiative generation of neutrino masses in the 3-3-1 scenario have been proposed in Refs.~\cite{Boucenna:2014ela,Okada:2015bxa,Fonseca:2016xsy,Machado:2018sfh}. 
These models, however, do not contain dark matter candidates. 
Recently, in Refs.~\cite{Kang:2019sab,Leite:2019grf} one-loop neutrino mass models including a dark matter candidate have been constructed 
by the extension of the minimal model based on the $SU(3)_C \otimes SU(3)_L \otimes U(1)_X \otimes U(1)_N$ gauge symmetry, so-called 3-3-1-1 models.  
In the minimal version of 3-3-1-1 models~\cite{Dong:2013wca,Dong:2014wsa,Dong:2015yra}, three right-handed neutrinos are introduced in order to 
realize the anomaly cancellation, and tiny masses of active neutrinos are obtained by the seesaw mechanism at tree level.
In addition, dark matter candidates are naturally obtained by the remnant discrete symmetry after the spontaneous breaking of the 3-3-1-1 gauge symmetry~\cite{Dong:2013wca,Dong:2015yra}. 
Scenarios with radiative generation of neutrino masses can also be considered in the framework of 3-3-1-1 models~\cite{Kang:2019sab,Leite:2019grf} by changing 
$U(1)_N$ charges for three right-handed neutrinos to avoid the tree level mass term while keeping the dark matter candidates.  
However, this model is a bit complicated, because many additional particles are further required for the anomaly cancellation and
for making all sterile neutrinos massive~\cite{Leite:2019grf}. 

In this paper, we construct a new model with the 3-3-1 gauge symmetry
in order to explain tiny neutrino masses, dark matter and generation of chiral fermions simultaneously. 
We take the minimal content of fermions required for the gauge anomaly cancellation, while we introduce an additional 
scalar $SU(3)_L$ triplet field to break the lepton number by the scalar self-interactions. 
Our model then induces ``Zee model'' like diagrams after the spontaneous breaking of $SU(3)_L \otimes U(1)_X \to SU(2)_L \otimes U(1)_Y$. 
It has been known that the neutrino oscillation data cannot be reproduced by the Zee model due to a too restricted structure of lepton flavor violating (LFV) interactions, see e.g.,~\cite{Hasegawa:2003by,He:2003ih}. 
On the contrary, our model includes additional sources of LFV interactions, so that we can explain current neutrino data. 
Another interesting feature of our model is that the appearance of an unbroken discrete $Z_2$ parity, which guarantees the stability of dark matter candidate, i.e., the lightest neutral particle with a $Z_2$-odd charge. 
This discrete symmetry arises as the remnant symmetry of the 3-3-1 gauge symmetry and a global $U(1)'$ symmetry\footnote{The appearance of such remnant unbroken $Z_2$ parity in 3-3-1 models has been pointed out in Refs.~\cite{Fregolente:2002nx,Hoang:2003vj}.}, where the  
latter is softly-broken and is introduced in order to avoid dangerous flavor changing neutral interactions between a SM fermion and an extra fermion. 
We find that the $Z_2$-odd scalar dark matter can explain the thermal relic abundance satisfying the current direct search results. 

This paper is organized as follows. 
In Sec.~\ref{sec:model}, we show the particle content in our model, and give the Higgs potential,  kinetic terms for scalar fields and Yukawa interaction terms. 
In Sec.~\ref{sec:neutrino}, we discuss the generation mechanism for neutrino masses and their mixings. 
Constraints from LFV decays of charged leptons are considered in Sec.~\ref{sec:lfv}. We show numerical results for the correlations between branching ratios of the LFV decays in the parameter sets satisfying the current neutrino data. 
Sec.~\ref{sec:pheno} is devoted for the discussion of the dark matter and the collider phenomenology. 
We conclude the article in Sec.~\ref{sec:conclusion}. 
In appendices, we present the mass formulae of physical scalar bosons (App.~\ref{sec:mass}) and those for 
the decay branching ratios for the LFV decays (App.~\ref{sec:app-lfv}).

\section{Model \label{sec:model}}

In this section, we define our model based on the 3-3-1 gauge symmetry.  
We first present the particle content and give expressions for component fields of $SU(3)_L$ (anti-)triplet fields. 
We then discuss the Higgs potential, kinetic terms for scalar fields and Yukawa interaction terms in the following subsections in order. 

\subsection{Particle content\label{sec:particle}}

\begin{table}[!h]
\begin{center}
\begin{tabular}{|c||ccccccccc|cccc|}\hline
          & \multicolumn{9}{c|}{Fermion}  &  \multicolumn{4}{c|}{Scalar} \\\hline\hline
 Fields   &  $L_L^i$         & $e_R^i$     & $E_R^i$   & $Q_L^a$         & $Q_L^{3}$  & $d_R^i$ & $B_R$        & $u_R^i$ & $U_R^a$   & $\Phi_1$ & $\Phi_2$ & $\Phi_3$   & $\Phi_\ell$ \\\hline\hline
$SU(3)_C$ & ${\bf 1}$        & ${\bf 1}$  & ${\bf 1}$ & ${\bf 3}$        & ${\bf 3}$  & ${\bf 3}$ & ${\bf 3}$  & ${\bf 3}$  & ${\bf 3}$  & ${\bf 1}$   & ${\bf 1}$  & ${\bf 1}$ & ${\bf 1}$   \\\hline
$SU(3)_L$ & ${\bf 3}$  & ${\bf 1}$  & ${\bf 1}$ & ${\bf \bar{3}}$  & ${\bf 3}$  & ${\bf 1}$ & ${\bf 1}$  & ${\bf 1}$  & ${\bf 1}$  & ${\bf 3}$ & ${\bf 3}$ & ${\bf 3}$ &${\bf 3}$   \\\hline
$U(1)_X$  & $-2/3$           &  $-1$      &  $-1$     & $1/3$              & $0$      &  $-1/3$   &  $-1/3$    &  $2/3$     &  $2/3$     & $1/3$ &  $-2/3$       &  $1/3$  & $4/3$               \\\hline\hline
$U(1)'$   & 0                &  $-q$      &  $2q$     & $0$              &  $0$       & $-q$      & $2q$       &  $q$    &  $-2q$        & $q$ &  $-q$       &  $-2q$  & $0$               \\\hline
\end{tabular}
\caption{Particle content and charge assignment under the gauge symmetry $SU(3)_C \otimes SU(3)_L \otimes U(1)_X$.  
The $U(1)'$ symmetry is a global, which is softly-broken by scalar interactions. 
The flavor indices $i$ and $a$ run over 1--3 and 1--2, respectively. 
 }
\label{particle}
\end{center}
\end{table}

The particle content and the charge assignment under the 3-3-1 gauge symmetry are shown in Table~\ref{particle}, in which 
the fermion content is taken to be minimal based on the requirement of the gauge anomaly cancellation. 
From the cancelation of the pure $SU(3)_L$ gauge anomaly, generation must be 3, number of the color, which 
can be regarded as the origin of the three-generation structure for leptons and quarks~\cite{Frampton:1992wt}. 
In addition to the gauge symmetry, we introduce a global $U(1)'$ symmetry, which is softly-broken.
This $U(1)'$ symmetry is imposed to avoid the dangerous flavor changing neutral current, while it maintains necessary scalar interaction terms in the potential to generate one-loop induced neutrino masses. 

The embedding scheme of  leptons and quarks into 3-3-1 multiplets is the same as that given in Ref.~\cite{Ozer:1995xi}, in which 
the electric charge $Q$ is defined by 
\begin{align}
Q = T_3 +Y,  ~~{\rm with}~~ Y = \frac{1}{\sqrt{3}}T_8 + Q_X, \label{charge}
\end{align} 
where $T_3$, $T_8$ and $Q_X$ are the third and the eighth components of the $SU(3)_L$ generator and the $U(1)_X$ charge, respectively. 
After the 3-3-1 symmetry is broken into the $SU(2)_L \otimes U(1)_Y$ symmetry, 
$Y$ in Eq.~(\ref{charge}) is identified with the weak hypercharge.  

The component fields of left-handed leptons and quarks are then determined as follows:
\begin{align}
L_L^i &= \begin{pmatrix}
\nu^i \\
e^i \\
E^i
\end{pmatrix}_L,~~
Q_L^a = \begin{pmatrix}
d^a \\
u^a \\
U^a
\end{pmatrix}_L,~~
Q_L^3 = \begin{pmatrix}
t \\
b \\
B
\end{pmatrix}_L, \label{lefth}
\end{align}
where the flavor indices $i$ and $a$ run over 1-3 and 1-2, respectively. 
In Eq.~(\ref{lefth}), component fields denoted as lowercase (uppercase) letters can be identified as SM (extra) fermions. 
We have three, two and one extra charged leptons, up-type quarks and down-type quark, respectively, 
whose electric charges are the same as the corresponding charged leptons and quarks in the SM. 
All these extra fermions have Dirac mass terms proportional to the Vacuum Expectation Value (VEV) which 
breaks the $SU(3)_L \otimes U(1)_X$ symmetry into the $SU(2)_L \otimes U(1)_Y$ symmetry. 

The scalar triplet fields are represented as 
\begin{align}
\Phi_{1} = \begin{pmatrix}
\phi_{1}^+ \\
\phi_{1}^0 \\
\eta_{1}^0
\end{pmatrix},~~
\Phi_2 = \begin{pmatrix}
\phi_2^0 \\
\phi_2^- \\
\eta_2^-
\end{pmatrix},~~
\Phi_{3} &= \begin{pmatrix}
\eta_{3}^+ \\
\eta_{3}^0 \\
\phi_{3}^0
\end{pmatrix},~~
\Phi_\ell = \begin{pmatrix}
\eta_\ell^{++} \\
\eta_\ell^+ \\
\phi_\ell^+
\end{pmatrix}. \label{eq:higgs}
\end{align}
In 3-3-1 gauge theories the Higgs sector contains at least three $SU(3)_L$-triplet scalar fields~\cite{Singer:1980sw,Valle:1983dk,Ozer:1995xi}, which give masses to the fermions except for the neutrinos after the electroweak symmetry breaking. 
In our model, we further introduce an additional triplet field $\Phi_\ell$ for the neutrino mass generation, which will be discussed in Sec.~\ref{sec:neutrino}. 

\subsection{Higgs potential \label{sec:potential}}

The most general Higgs potential is given by 
\begin{align}
V &=  \sum_{i=1,4} m_i^2 |\Phi_i|^2  + (m_{13}^2 \Phi_1^\dagger  \Phi_3   + \epsilon_{\alpha\beta\gamma} \, \mu \Phi_1^\alpha \,\Phi_2^\beta \,\Phi_3^\gamma  +  \text{h.c.}) \notag\\
& + \sum_{i=1,4}\lambda_i|\Phi_i|^4 + \sum_{i,j=1,4}^{j>i} \left(\lambda_{ij}|\Phi_i|^2|\Phi_j|^2  + \rho_{ij}|\Phi_i^\dagger \Phi_j|^2 \right) \notag\\
& +\xi_1 (\Phi_1^\dagger \Phi_2)(\Phi_3^\dagger \Phi_\ell) + \xi_2 (\Phi_2^\dagger \Phi_3)(\Phi_\ell^\dagger \Phi_1) + \text{h.c.}, \label{eq:pot}
\end{align}
where $\Phi_4 = \Phi_\ell$. 
The $m_{13}^2$ and $\mu$ terms softly break the $U(1)'$ symmetry, by which the appearance of an additional Nambu Goldstone (NG) boson is avoided. 
The parameters $m_{13}^2$, $\mu$, $\xi_1$ and $\xi_2$ are complex in general, and these complex phases cannot be simultaneously taken to be zero 
by phase redefinitions of the scalar fields. For simplicity, we take these parameters to real. 

The VEVs of the Higgs triplet fields can generally be taken as 
\begin{align}
\langle\Phi_1\rangle = \frac{1}{\sqrt{2}}\begin{pmatrix}
0 \\
v_1 \\
0
\end{pmatrix},~~
\langle\Phi_2\rangle = \frac{1}{\sqrt{2}}\begin{pmatrix}
v_2 \\
0 \\
0
\end{pmatrix}, ~~
\langle\Phi_3\rangle &= \frac{1}{\sqrt{2}}\begin{pmatrix}
0 \\
v' \\
V
\end{pmatrix}.  \label{eq:vev}
\end{align}
The VEV of the third component of $\Phi_1$ can be taken to zero without any loss of generality by using 
the field rotation of $\Phi_1$ and $\Phi_3$. 
The VEV $v'$ causes phenomenologically dangerous mixing between the SM fermions and the extra ones. 
Therefore, we arrange $\mu\neq 0$ and $m_{13}^2 = 0$ so as to have a remnant $Z_2$ symmetry (denoting it by $\tilde{Z}_2$), by which $v'= 0$ 
is guaranteed and such a dangerous mixing can be avoided. For details, see Appendix~\ref{sec:stationary}. 
The charge of the $\tilde{Z}_2$ symmetry can be defined as $(-1)^{|Q'|/q}$ with $Q'$ being the $U(1)'$ charge, 
by which $\Phi_{1,2}$ and all the SM right-handed fermions are assigned to be odd, while the other fields are even.  
The $\tilde{Z}_2$ symmetry is spontaneously broken by the VEVs $v_1$ and $v_2$, so that domain walls would appear in the early Universe~\cite{Zeldovich:1974uw,Kibble:1976sj}. 
We will briefly comment on this issue at the end of this subsection. 

In the following, we assume that $V \gg v_1,v_2$. 
Under the setup with $v' = 0$, $SU(3)_L \otimes U(1)_X$ is spontaneously broken into 
$SU(2)_L \otimes U(1)_Y$ by the VEV $V$ at higher energy scales than the electroweak scale. 
Then, the $SU(2)_L \otimes U(1)_Y$ symmetry is  broken down to $U(1)_{\rm em}$ by $v_1$ and $v_2$ 
at the electroweak scale. 
The Fermi constant $G_F$ is reproduced by $G_F = (\sqrt{2}v^2)^{-1}$ with $v \equiv \sqrt{v_1^2 + v_2^2}$. 
For the later convenience, we introduce $\tan\beta = v_2/v_1$ as the analogue of two Higgs doublet models (THDMs).

%% \begin{table}[t]
%% \begin{center}
%% \begin{tabular}{|c||ccccccccc|cc|}\hline
%%  Fields             &  $L_L^i$        & $e_R^i$     & $E_R^i$   & $Q_L^a$         & $Q_L^{3}$     & $d_R^i$ & $B_R$        & $u_R^i$ & $U_R^a$   & $\Phi_{1,2}$ & $\Phi_{3,\ell}$  \\\hline
%% $Z_2^{\text{rem}}$    &$(+,+,-)^T$      &  $+$      &  $-$       & $(+,+,-)^T$     &  $(+,+,-)^T$ & $+$      & $-$       &  $+$    &  $-$        & $(+,+,-)^T$ &  $(-,-,+)^T$       \\\hline
%% \end{tabular}
%% \caption{$Z_2^{\text{rem}}$ charges for each component field. }
%% \label{particle2}
%% \end{center}
%% \end{table}

\begin{table}[t]
\begin{center}
\begin{tabular}{|c||cccc|ccc|cccc|}\hline
          & \multicolumn{4}{c|}{Fermions}  &  \multicolumn{3}{c|}{Scalar bosons} &  \multicolumn{4}{c|}{Gauge bosons} \\\hline\hline
$Z_2^{\text{rem}}$-even fields    &  $e^i$  & $u^i$     & $d^i$  & $\nu_L^i$       & $\phi_{1,2,3}^0$      & $\phi_{1,2,\ell}^\pm$  &                      & $\gamma^\mu$     & $Z^\mu$ & $Z^{\prime \mu}$ &  $W^\mu$ \\\hline
$Z_2^{\text{rem}}$-odd fields     &  $E^i$  & $U^a$     & $B$    &                 & $\eta_{1,3}^0$       & $\eta_{1,2,\ell}^\pm$  & $\eta_{\ell}^{\pm\pm}$ & $W^{\prime \mu}$  & $Y^\mu$ &                 &\\\hline
\end{tabular}
\caption{$Z_2^{\text{rem}}$ charges of particles. Definitions for the gauge bosons are given in Sec.~\ref{sec:kin}. }
\label{particle2}
\end{center}
\end{table}

After the spontaneous breakdown of the 3-3-1 gauge symmetry and the $\tilde{Z}_2$ symmetry, 
another remnant $Z_2$ symmetry, let us denote it as $Z_2^{\rm rem}$, appears, whose charge can be defined as $(-1)^{2s + 2\sqrt{3}T_8 + |Q'|/q}$ with $s$ being the spin of the particle. 
In Table~\ref{particle2}, we show the $Z_2^{\rm rem}$ charges for each particle, where the charges of gauge bosons can be determined from their structures of interactions, see Sec.~\ref{sec:kin}. 
Because the $Z_2^{\rm rem}$ symmetry is unbroken, the lightest neutral $Z_2^{\rm rem}$-odd particle can be a candidate of dark matter. 
We will discuss dark matter physics in Sec.~\ref{sec:pheno}. 

Because of the $Z_2^{\rm rem}$ symmetry, we can classify the physical scalar fields into the $Z_2^{\rm rem}$-even and $Z_2^{\rm rem}$-odd ones as follows. 
In the $Z_2^{\rm rem}$-even sector, we have two-pairs of singly-charged scalar bosons $H^\pm$ and $\hat{H}^\pm$, 
one CP-odd Higgs boson $A$ and three CP-even Higgs bosons $H_i$ ($i = 1,2,3$). 
The discovered Higgs boson with a mass of about 125 GeV can be identified with the $H_1$ state. 
On the other hand in the $Z_2^{\rm rem}$-odd sector, we have one-pair of doubly-charged scalar bosons $\eta_\ell^{\pm\pm}$, 
two-pairs of singly-charged scalar bosons $\eta^\pm$ and $\hat{\eta}^\pm$, and one complex neutral scalar boson $\eta^0$. 
The other eight scalar states are the NG bosons which are absorbed into the longitudinal components of the massive gauge bosons ($W_\mu$, $W'_\mu$, $Y_\mu$, $Z_\mu$ and $Z'_\mu$), see Sec.~\ref{sec:kin}. 
In Appendix~\ref{sec:mass}, we explicitly show the relation between the mass eigenstates and the weak eigenbasis of the scalar states and their mass formulae. 

Let us discuss the effective theory of our 3-3-1 model in the large VEV limit $V \gg v$ with $V\mu \equiv  M^2$. 
In this case, the masses of $H^\pm$, $A$ and $H_2$ are determined by the $M$ parameter\footnote{This parameter plays the similar role to a soft-breaking $Z_2$ parameter in THDMs. }, while 
that of $H_1$ is determined by $v$. 
On the other hand, all the other physical Higgs bosons are decoupled from the theory, as their masses are determined by $V$. 
Therefore, the scalar sector effectively coincides with a THDM with a special flavor structure which cannot be realized in THDMs with a softly-broken $Z_2$ symmetry, see Sec.~\ref{sec:yukawa}. 
Similar to the usual THDMs, we can define the decoupling limit by $M \gg v$, where only the SM-like Higgs boson $H_1$ remains at the scale $v$. 
We can also define the so-called alignment limit, where the SM-like Higgs boson couplings with the SM gauge bosons and fermions
become the same values as those of the SM Higgs boson at tree level. 
This alignment limit can be taken by choosing potential parameters such that the (1,2) element of the mass matrix of the CP-even Higgs bosons given in Eq.~(\ref{eq:cp-even}) is zero. 
Therefore, our 3-3-1 model provides another important example that predicts the THDM as the low energy effective theory other than the minimal supersymmetric extension of the SM~\cite{Haber:1984rc} and composite Higgs models~\cite{Mrazek:2011iu,DeCurtis:2018zvh,DeCurtis:2018iqd}.

As mentioned in the above, our model potentially has the domain wall problem. 
It has been known that the energy density of domain walls is only suppressed by the inverse of the radius of the Universe, which is much slower 
than the dilution of the energy density for ordinary matter and radiation. 
Therefore, the existence of domain walls could significantly change the history of the Universe. 
In Refs.~\cite{Dvali:1995cc,Dvali:1996zr,Preskill:1991kd,Riva:2010jm}, solutions for the domain wall problem have been discussed. 
According to Ref.~\cite{Dvali:1995cc}, if a discrete symmetry which is spontaneously broken by Higgs VEVs (in our model, this corresponds to the electroweak symmetry breaking VEV $v$)
is not restored at high temperature, the domain wall problem might not arise. 
Such situation can happen if finite temperature effects which are proportional to $T^2$ on a negative mass squared term are also negative~\cite{Weinberg:1974hy}. 
In the SM, this does not happen, because there is only one scalar quartic coupling. 
Such quartic coupling gives a positive effect of finite temperature on the negative mass squared term, so that 
the broken symmetry at zero temperature is restored at high temperature as it is seen in the usual thermal history of the Universe. 
On the other hand, if we consider models with multi-scalar fields as in our model, 
this is not always the case, because some combinations of scalar quartic parameters can be taken to negative so as to realize the symmetry non-restoration scenario. 
Thus, we might be able to avoid the domain wall problem. Clearly, more dedicated discussions for this solution have to be done in order to ensure its justification, 
which is beyond the scope of this paper. 

\subsection{Kinetic terms for scalar fields\label{sec:kin}}

Kinetic terms for the scalar triplet fields are expressed as 
\begin{align}
{\cal L}_{\rm kin} = \sum_{i = 1,4}|D_\mu \Phi_i|^2, 
\end{align}
where $\Phi_4 = \Phi_\ell$. The covariant derivative $D_\mu$ for $SU(3)_L$ triplet fields is given by 
\begin{align}
D_\mu = \partial_\mu - igA_\mu -ig_X^{}Q_XX_\mu, 
\end{align}
with $g$ and $g_X$ being the $SU(3)_L$ and $U(1)_X$ gauge couplings, respectively. 
The $SU(3)_L$ gauge boson $A_\mu$ is expressed by the $3\times 3$ matrix form as:
\begin{align}
A^\mu \equiv A^{A\mu} T^A =  
\begin{pmatrix}
\frac{1}{2}(A_3^\mu + \frac{A_8^\mu}{\sqrt{3}}) & \frac{A_{12}^{\mu}}{\sqrt{2}} & \frac{A_{45}^\mu}{\sqrt{2}}\\
\frac{A_{12}^{*\mu}}{\sqrt{2}} & \frac{1}{2}(-A_3^\mu + \frac{A_8^\mu}{\sqrt{3}}) & \frac{A_{67}^{\mu}}{\sqrt{2}} \\
\frac{A_{45}^{*\mu}}{\sqrt{2}} & \frac{A_{67}^{*\mu}}{\sqrt{2}} & -\frac{A_8^\mu}{\sqrt{3}}
\end{pmatrix},\quad A = 1,\dots, 8, 
\label{vmat}
\end{align}
where we introduced $A_{ij}^\mu \equiv (A_i^\mu - iA_j^\mu)/\sqrt{2}$ with $(i,j) = (1,2)$, (4,5) and (6,7). 
We can identify $W^\mu \equiv A_{12}^\mu$ and $W^{\prime\mu} \equiv A_{45}^\mu$ with the SM W boson and the additional charged gauge boson, respectively, while 
$Y^\mu \equiv A_{67}^\mu$ with a neutral complex gauge boson. 
Their masses are given by the VEVs of the Higgs triplet fields as follows: 
\begin{align}
m_W = \frac{g}{2}v,\quad 
m_{W'} = \frac{g}{2}\sqrt{v^2c_\beta^2 + V^2}, \quad
m_{Y} = \frac{g}{2}\sqrt{v^2s_\beta^2 + V^2}, 
\end{align}
where $s_\theta = \sin\theta$ and $c_\theta = \cos\theta$. 
Notice here that the gauge bosons $W'$ and $Y$ appear in the (1,3) and (2,3) elements of the matrix given in Eq.~(\ref{vmat}), so that 
they interact with one $Z_2^{\rm rem}$-even and one $Z_2^{\rm rem}$-odd fermions or scalar bosons in their trilinear interactions. 
Thus, $W'$ and $Y$ can be identified with the $Z_2^{\rm rem}$-odd particles. 
The other gauge bosons can be identified with the $Z_2^{\rm rem}$-even particles, because they interact with two $Z_2^{\rm rem}$-even or two $Z_2^{\rm rem}$-odd fermions (scalar bosons). 
This property was already summarized in Table~\ref{particle2}. 

In addition to these complex states, there are three real neutral gauge bosons, where one of them can be identified with the massless photon $\gamma^\mu$. 
We can define the basis where $\gamma^\mu$ is separated from the other two massive states ($\tilde{Z}^\mu$ and $\tilde{Z}^{\prime \mu}$) as follows: 
\begin{align}
\begin{pmatrix}
A_3^\mu \\
A_8^\mu \\
X^\mu 
\end{pmatrix}
=R_V
\begin{pmatrix}
\gamma^\mu \\
\tilde{Z}^\mu \\
\tilde{Z}^{\prime \mu} 
\end{pmatrix}, \quad 
R_V = 
\left(
\begin{array}{ccc}
 \frac{\sqrt{3} g_X^{}}{\sqrt{3 g^2+4
   g_X^2}} & \frac{1}{2} & -\frac{3g}{2
   \sqrt{3 g^2+4 g_X^2}} \\
 \frac{g_X^{}}{\sqrt{3 g^2+4 g_X^2}} &
   -\frac{\sqrt{3}}{2} & -\frac{\sqrt{3} g}{2 \sqrt{3
   g^2+4 g_X^2}} \\
 \frac{\sqrt{3} g}{\sqrt{3 g^2 + 4 g_X^2}} & 0 &
   \frac{2g_X}{\sqrt{3 g^2 + 4g_X^2}}
\end{array}
\right). 
\end{align}
The $\tilde{Z}$ and $\tilde{Z}'$ states can be mixed. Their mass matrix is given by 
\begin{align}
{\cal M}_V^2 &= \frac{V^2}{4}
\begin{pmatrix}
g^2(1 +  c_\beta^2\epsilon) & \frac{g}{3}\sqrt{3g^2 + 4g_X^2}(1 -  c_\beta^2\epsilon) \\
 \frac{g}{3}\sqrt{3g^2 + 4g_X^2}(1- c_\beta^2\epsilon) &  \frac{1}{9}(3g^2+4g_X^2)[1 + (1+3s_\beta^2)\epsilon ]
\end{pmatrix}, 
\end{align}
with $\epsilon = v^2/V^2$. 
In the large $V$ limit, the mass eigenvalues are expressed as 
\begin{align}
m_Z^2 &= \frac{g^2v^2}{4}\left[\frac{3g^2+4 g_X^2}{3 g^2+g_X^2} + {\cal O}(\epsilon)\right], \quad m_{Z'}^2 = \frac{V^2}{9}\left[3g^2 + g_X^2 + {\cal O}(\epsilon)\right]. 
\end{align}
From the expression of $m_Z^2$, we see that the weak mixing angle $\theta_W$ can be identified with
\begin{align}
c_W \equiv \cos\theta_W = \sqrt{\frac{3 g^2+g_X^2}{3g^2+4 g_X^2}}. \label{eq:theta}
\end{align}
We then have the same expression of $m_Z$ as that of the mass of the SM $Z$ boson. The $U(1)_{\rm em}$ coupling $e$ is consistently given by $e = g\sin\theta_W$ as that in the SM. 

The electroweak rho parameter can be expressed at tree level by using the definition of $\theta_W$ given in Eq.~(\ref{eq:theta}), 
\begin{align}
\rho = \frac{m_W^2}{m_Z^2 c_W^2} = 1 + {\cal O}(\epsilon). \label{eq:rho}
\end{align}
In order to satisfy $|\rho -1| \leq 10^{-3}$, $V$ has to be taken to be larger than around 8 TeV. 
We note that the calculation of one-loop corrections to the rho parameter is different from that in models with $\rho =1$ at tree level, e.g., the SM. 
In models with $\rho \neq 1$ at tree level, the electroweak sector is described by four input parameters which can be chosen to be
$\alpha_{\rm em}$, $m_Z$, $G_F$ and $\delta \rho$ with the last one being the deviation of the rho parameter from unity. 
This means that the radiative correction to the rho parameter cannot be a prediction, because the additional parameter $\delta \rho$ provides  
an additional counterterm by which loop corrections to the rho parameter can be absorbed by imposing a renormalization condition. 
The similar situation can also happen in models with higher isospin scalar multiplets with a non-vanishing VEV such as $SU(2)_L$ triplet scalar fields~\cite{Blank:1997qa,Kanemura:2012rs}.

\subsection{Yukawa interactions\label{sec:yukawa}}

Thanks to the global $U(1)'$ symmetry, Yukawa interaction terms for the SM right-handed fermions and those for the extra right-handed fermions
are separately given as follows: 
\begin{align}
\mathcal{L}_Y &=
\epsilon_{\alpha\beta\gamma}f_{ij} (\overline{L_L^{ci}})_\alpha (L_L^j)_\beta (\Phi_\ell)_\gamma + y_e^{ij}\bar{L}_L^i \Phi_1 e_R^j    \notag\\
& +  y_{d2}^{ai}\bar{Q}_L^a\Phi_2^* d_R^i  +  y_{d1}^{i}\bar{Q}_L^3\Phi_1 d_R^i  
 +  y_{u1}^{ai}\bar{Q}_L^a \Phi_1^* u_R^i +  y_{u2}^{ai}\bar{Q}_L^3\Phi_2 u_R^i  \notag\\
& +  y_E^{ij}\bar{L}_L^i \Phi_3 E_R^j  + y_D\bar{Q}_L^3\Phi_3 B_R  + y_U^{ab}\bar{Q}_L^{a}\Phi_3^* U_R^b + \text{h.c.}, \label{yuk}
\end{align}
where $\epsilon^{\alpha\beta\gamma}$ is the complete antisymmetric tensor with $\alpha$, $\beta$ and $\gamma$ being the indices for the $SU(3)_L$ triplet. 
Because of $\epsilon^{\alpha\beta\gamma}$, the complex $3\times 3$ matrix $f_{ij}$ has also to be antisymmetric. 

From the structure of the VEVs and the Yukawa interactions given in Eqs. (\ref{eq:vev}) and (\ref{yuk}), respectively, 
the mass matrices for the SM fermions ($M_f$ with $f =u,d,e$) and those for the extra fermions ($M_F$ with $F =U,D,E$) are given by: 
\begin{align}
M_f = \frac{v}{\sqrt{2}}Y_f, \quad M_F = \frac{V}{\sqrt{2}}y_F^{}, 
\end{align}
where 
\begin{align}
Y_e = y_e c_\beta,~
Y_d = \begin{pmatrix}
y_{d2}^{11} s_\beta & y_{d2}^{12} s_\beta & y_{d2}^{13} s_\beta \\
y_{d2}^{21} s_\beta & y_{d2}^{22} s_\beta & y_{d2}^{23} s_\beta \\
y_{d1}^{1} c_\beta & y_{d1}^{2} c_\beta & y_{d1}^{3} c_\beta 
\end{pmatrix},~
Y_u = \begin{pmatrix}
y_{u1}^{11} c_\beta & y_{u1}^{12} c_\beta & y_{u1}^{13} c_\beta \\
y_{u1}^{21} c_\beta & y_{u1}^{22} c_\beta & y_{u1}^{23} c_\beta \\
y_{u2}^{1} s_\beta & y_{u2}^{2} s_\beta & y_{u2}^{3} s_\beta  
\end{pmatrix}.  \label{eq:mass}
\end{align}
We note that $M_E$, $M_U$ and $M_D$ are respectively the $3\times 3$, the $2\times 2$ and the $1\times 1$ matrices. 
They are diagonalized by bi-unitary transformations: 
\begin{align}
f_{L,R}^{} = V_{L,R}^f f_{L,R}',\quad
F_{L,R}    = V_{L,R}^F F_{L,R}'.  \label{eq:biunitary}
\end{align}
As we can see in Eq.~(\ref{eq:mass}), the mass matrix for the SM charged leptons has the same form as in the SM; i.e., 
only one of the Higgs fields gives their masses. 
On the other hand, the mass matrices for the up-type and the down-type SM quarks are given by two VEVs of the Higgs fields; i.e., $3\times 3$ matrices are composed of the two independent Yukawa matrices. 
This structure predicts characteristic flavor-dependent Higgs-boson couplings to quarks~\cite{Okada:2016whh}
in the THDM which is effectively deduced after the $SU(3)_L \otimes U(1)_X$ symmetry breaking as discussed in Sec.~\ref{sec:potential}. 

The $f_{ij}$ terms in Eq.~(\ref{yuk}) do not contribute to the masses of charged leptons, but they play an important role for the neutrino mass generation. 
From these interaction terms, we can assign two units of the lepton number to the Higgs triplet field $\Phi_\ell$. 
This lepton number is explicitly broken by the $\xi_1$ and $\xi_2$ terms in the Higgs potential given in Eq.~(\ref{eq:pot}), and they turn out to be the source of the Majorana masses for the active neutrinos. 
We will discuss the neutrino mass generation in Sec.~\ref{sec:neutrino}.

\section{Neutrino masses \label{sec:neutrino}}

\begin{figure}[t]
\begin{center}
\includegraphics[width=180mm]{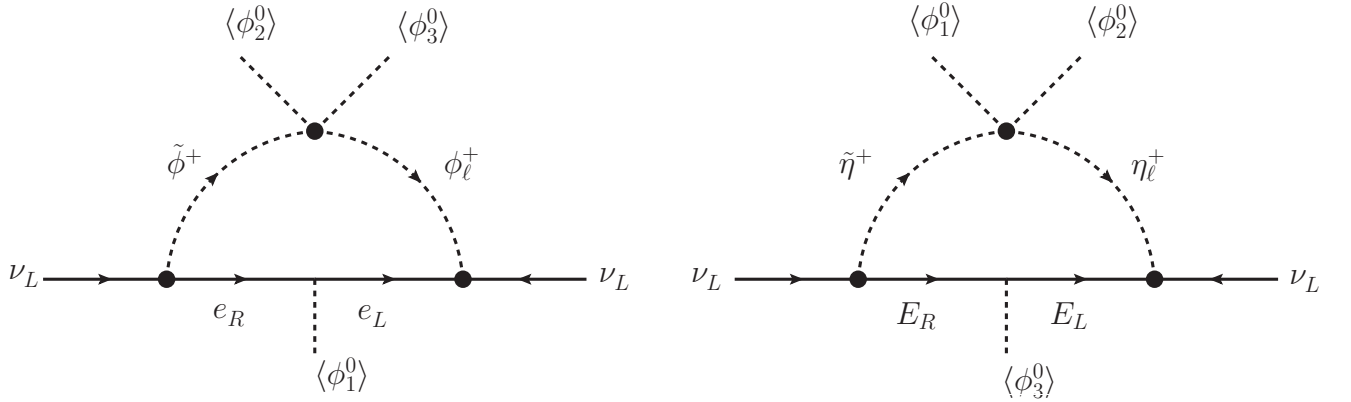}
\caption{One-loop diagrams for the neutrino mass generation.
Charged scalar fields are written in the basis where the NG boson fields are separated, see Appendix~\ref{sec:mass}. }
\label{fig1}
\end{center}
\end{figure}

Majorana neutrino masses are generated at one-loop level as shown in Fig.~\ref{fig1}, in which 
the $Z_2^{\rm rem}$-even and $Z_2^{\rm rem}$-odd particles run in the loop in the left and right diagram, respectively. 
%% We note that these diagrams effectively induce the following dimension 6 operator: 
%% \begin{align}
%% {\cal L}_{\rm eff} = \frac{c_{ij}}{M_6^2}\bar{L}_L^{ci}L_L^j\Phi_1\Phi_3\Phi_2^*, \label{eq:dim6}
%% \end{align} 
%% where $M_6$ is a typical mass scale of new particles in the loop. 
%% This operator is allowed by the 3-3-1 gauge symmetry and the global $U(1)'$ symmetry, while the dimension 5 operator $(\bar{L}_L^c \Phi_2^*)(\Phi_2^\dagger L_L)$
%% is forbidden by the $U(1)'$ symmetry. 
%% Thus, we have three insertions of the VEVs to obtain Majorana mass terms as shown in Fig.~\ref{fig1}, which is consistent with the view point of the above effective field approach. 
%
We here give relevant interaction terms among physical charged scalar bosons and fermions in their mass eigenbases for the calculation of one-loop induced neutrino masses; 
\begin{align}
&{\cal L}_{\rm int} = \frac{\sqrt{2}}{V} \frac{s_\beta}{\sqrt{1 + \frac{v^2}{V^2}s_\beta^2}}  \bar{\nu}_L' (W M_E^{\rm diag})E_R' (c_{\theta_\eta}\eta^+ - s_{\theta_\eta}\hat{\eta}^+)
-\frac{\sqrt{2}\tan\beta}{v}\bar{\nu}_L' M_e^{\rm diag} e_R' (c_{\theta_H}H^+ - s_{\theta_H}\hat{H}^+)\notag\\
& -2[\bar{\nu}_L^{\prime c} F e_L' (s_{\theta_H} H^+ + c_{\theta_H} \hat{H}^+) 
-\bar{\nu}_L^{\prime c} F W E_L' (s_{\theta_\eta} \eta^+ + c_{\theta_\eta} \hat{\eta}^+) 
-\bar{E}_L^{\prime c} W^TFe_L'\eta_\ell^{++} ] + \text{h.c.} , \label{eq:int2}
\end{align}
where $F \equiv V_L^T f V_L$, $W = (V_L^e)^\dagger V_L^E$, and $\theta_H$ and $\theta_\eta$ being the mixing angles of charged scalar fields, see Eq.~(\ref{mixing1}) and (\ref{mixing2}). 
In the above expression, $M_e^{\rm diag}$ and $M_E^{\rm diag}$ are the diagonalized mass matrices for the SM charged leptons and those for the extra leptons, respectively. 
The dashed fields are related to the original one by Eq.~(\ref{eq:biunitary}). 
It is important to mention here that the appearance of the $W$ matrix which is the $3\times 3$ unitary matrix similar to the Cabibbo-Kobayashi-Maskawa matrix plays a crucial role 
to reproduce the current neutrino oscillation data as it will be clarified below. 

The total contribution to the Majorana neutrino masses is then expressed as 
\begin{align}
{\cal M}_\nu = ({\cal M}_\nu^e + {\cal M}_\nu^{E}) + ({\cal M}_\nu^e + {\cal M}_\nu^{E})^T, \label{eq:numass}
\end{align}
where ${\cal M}_\nu^e$ and ${\cal M}_\nu^E$ represent the contribution from the left and right diagram, respectively. 
They are calculated as 
\begin{align}
({\cal M}_\nu^e)_{ij} &=  \frac{C_e}{v} F_{ij}m_j^2  , \quad ({\cal M}_\nu^E)_{ij} =  \frac{C_E}{V} (FW)_{ik}M_k^2 G_k (W^\dagger)_{kj} ,  \label{eq:mnu}
\end{align}
where $m_i \equiv (M_e^{\rm diag})_{ii}$ and $M_i \equiv (M_E^{\rm diag})_{ii}$, and 
\begin{align}
G_k &= \frac{1}{2}\ln \frac{m_{\eta^\pm}^2}{m_{\hat{\eta}^\pm}^2} + \frac{M_k^2 + m_{\hat{\eta}^\pm}^2}{M_k^2 - m_{\hat{\eta}^\pm}^2}\ln \frac{m_{\hat{\eta}^\pm}^2}{M_k^2}
- \frac{M_k^2 + m_{\eta^\pm}^2}{M_k^2 - m_{\eta^\pm}^2}\ln \frac{m_{\eta^\pm}^2}{M_k^2}. 
\end{align}
The flavor independent coefficients $C_e$ and $C_E$ are given by 
\begin{align}
C_e    & = \frac{\sqrt{2}}{8\pi^2}\tan\beta c_{\theta_H} s_{\theta_H}\ln \frac{m_{\hat{H}^\pm}^2}{m_{H^\pm}^2},  \quad 
C_E     = \frac{\sqrt{2}}{8\pi^2}\frac{s_\beta c_{\theta_\eta}s_{\theta_\eta}}{\sqrt{1 + \frac{v^2}{V^2}s_\beta^2}}. 
\end{align}
The mass matrix given in Eq.~(\ref{eq:numass}) is diagonalized by introducing the Pontecorvo-Maki-Nakagawa-Sakata matrix $U_{\rm PMNS}$ as; 
\begin{align}
U^T_{\rm PMNS}\, {\cal M}_\nu\, U_{\rm PMNS} = \text{diag}(m_\nu^1,m_\nu^2,m_\nu^3), 
\end{align}
where $m_{\nu}^i$ $(i=1,2,3)$ are mass eigenvalues for neutrinos, and 
\begin{align}
U_{\rm PMNS} = 
\begin{pmatrix}
1 & 0 & 0 \\ 
0 & c_{23} & s_{23} \\
0 & -s_{23} & c_{23}
\end{pmatrix}
\begin{pmatrix}
c_{13} & 0 & s_{13}e^{-i\delta_{\rm CP}} \\
0          & 1 & 0 \\
-s_{13}e^{i\delta_{\rm CP}} & 0 & c_{13}
\end{pmatrix}
\begin{pmatrix}
c_{12} & s_{12} & 0 \\
-s_{12} & c_{12} & 0 \\
0          & 0 & 1 
\end{pmatrix}, 
\end{align}
with $s_{ij} = \sin\theta_{ij}$, $c_{ij} = \cos\theta_{ij}$ and $\delta_{\rm CP}$ being the CP phase. 
We consider both the cases for the orders of the neutrino masses; i.e., the normal hierarchy ($|m_\nu^1| < |m_\nu^2| < |m_\nu^3|$) and the inverted hierarchy ($|m_\nu^3| < |m_\nu^1| < |m_\nu^2|$). 

The flavor structure of $M_\nu^e$ is the same as that of the Zee model. 
It has been known that the Zee model cannot explain the current neutrino oscillation data because of the too restricted structure of flavor violating couplings which 
only arise from the antisymmetric $3\times 3$ $F$ matrix, see e.g.,~\cite{Hasegawa:2003by,He:2003ih}\footnote{If we take the general Yukawa interactions for leptons, then there is 
a corner of parameter space which can satisfy the current neutrino results~\cite{Herrero-Garcia:2017xdu,Nomura:2019dhw,Babu:2019mfe}.}. 
Although the flavor structure of ${\cal M}_\nu^E$ also takes the similar form to that of ${\cal M}_\nu^e$, 
another flavor violating source in the matrix $W$ is inserted into the mass matrix. 
Consequently, ${\cal M}_\nu^E$ has a different flavor mixing pattern.  
Hence, it can explain the current neutrino data. 
We note that such an additional flavor violating source vanishes if we take the masses of the extra leptons degenerate; 
i.e., $M_1 = M_2 = M_3$, by which 
$M_k^2$ and $G_k$ in Eq.~(\ref{eq:mnu}) commute with $W^\dagger$ and then the effect of the $W$ matrix disappears by the unitarity. 
We also note that the contribution from ${\cal M}_\nu^E$ is typically much larger than that from ${\cal M}_\nu^e$, because the latter is proportional to the mass squared of the SM charged leptons.  
Therefore, the neutrino masses and the mixings are determined essentially only by the contribution from ${\cal M}_\nu^E$. 
We thus switch off the contribution from ${\cal M}_\nu^e$ hereafter for simplicity, 
which can be realized by taking $C_e = 0$ or equivalently $\theta_H = 0$. 

\section{Lepton Flavor Violation\label{sec:lfv}}

\begin{figure}[t]
\begin{center}
\includegraphics[width=160mm]{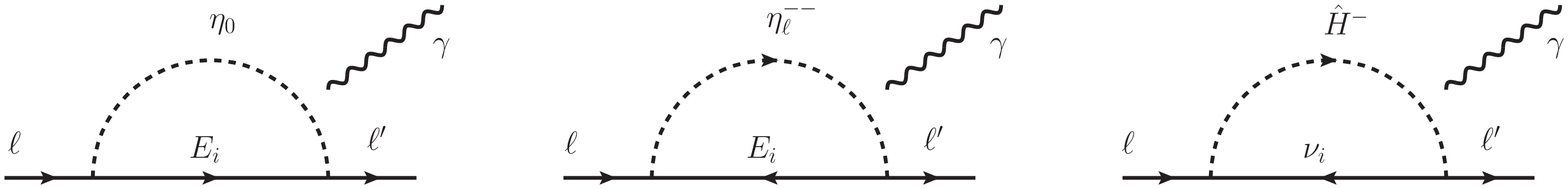}
\caption{Diagrams for $\ell \to \ell' \gamma$ processes. }
\label{fig2}
\end{center}
%\end{figure}
%
%\begin{figure}[t]
\begin{center}
\includegraphics[width=160mm]{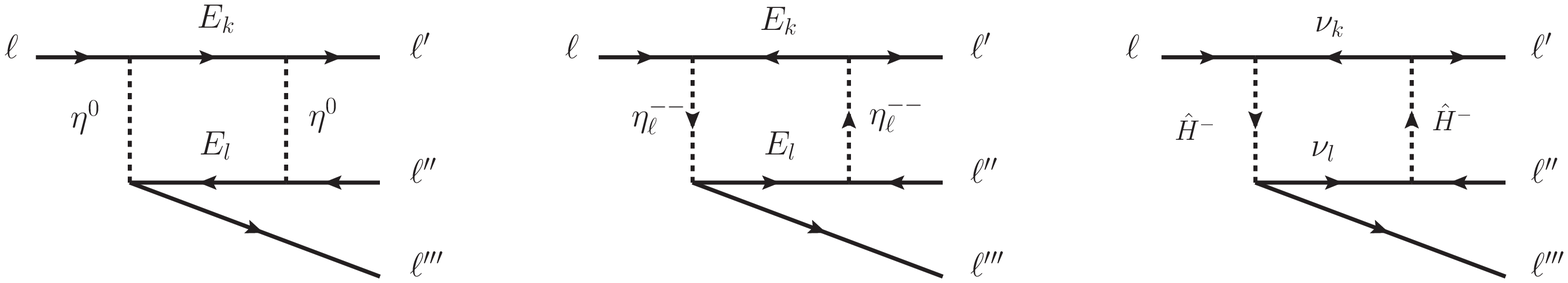}
\caption{Box diagrams for $\ell \to \ell'\ell''\ell'''$ processes. }
\label{fig3}
\end{center}
\end{figure}

In our model, new particles can contribute to LFV decays of the charged leptons. 
In this section, we discuss the constraints from $\ell \to \ell' \gamma$ and $\ell \to \ell'\ell''\ell'''$ types of LFV decays in the parameter sets which satisfy the current neutrino oscillation data. 

Diagrams of the $\ell \to \ell' \gamma$ processes are shown in Fig.~\ref{fig2}. 
The branching ratios of these processes are calculated as 
\begin{align}
{\cal B}(\ell_i \to \ell_j \gamma) \simeq 
\frac{48\pi^3 \alpha_{\rm em} C_{ij}}{G_F^2} \left[|(a_R)_{ij}|^2 +   |(a_L)_{ij}|^2\right],  \label{eq:lfv1}
\end{align}
where $(\ell_1,\ell_2,\ell_3) = (e,\mu,\tau)$ and $(C_{21},C_{31},C_{32} )= (1,0.1784,0.1736)$. 
In the above expression, the detailed formulae for $a_L^{}$ and $a_R^{}$ are presented in Appendix~\ref{sec:app-lfv}. 
From the first diagram in Fig.~\ref{fig2}, the structure of the $W$ matrix is constrained. 
Since $W$ is a unitary matrix, we cannot simply take small values for each component. 
We can instead take the small mass difference among $E_i$ 
in order to suppress the contribution from the first diagram. 
On the other hand, the magnitude of the contributions from the second and the third diagrams can be easily suppressed by taking small values for the $F$ matrix elements. 
Typically, $|F_{ij}|\lesssim 10^{-3}$ is required from the constraint by the $\mu \to e \gamma$ data. 

For the $\ell \to \ell'\ell''\ell'''$ processes, the branching ratio of $\mu \to 3e$ is most strongly constrained by the data among the six possible processes of this type. 
We thus concentrate on the constraint from the $\mu \to 3e$ data. 
In this case, there are penguin type diagrams with the photon and the Z boson exchanges and the box diagrams shown in Fig.~\ref{fig3}. 
We confirm that the contribution from the box diagrams to the branching ratio is typically eight orders of the magnitude smaller than the branching ratio of $\mu \to e\gamma$ in our scenario, so 
that we can safely neglect these contributions. 
In addition, it is usually the case that the contribution from the penguin diagram with the Z boson exchange is negligibly smaller than that of the photon exchange~\cite{Hisano:1995cp}\footnote{
If there are new particles with large isospin charges which contribute to the effective $\bar{\ell} \ell' Z$ vertex, 
then the Z penguin diagram could be important as well as the photon one.  This is, however, not the case in our model. }.
Therefore, the dominant contribution arises from the penguin diagrams with the photon exchange which can be obtained by attaching the electron-positron line to the photon in the diagrams shown in Fig.~\ref{fig2}. 
The branching ratio of $\mu \to 3e$ is expressed as 
\begin{align}
&{\cal B}(\mu \to ee\bar{e}) \notag\\
& \simeq  \frac{6\alpha_{\rm em}^2}{G_F^2}
\Bigg[\frac{2}{3}(|a_L|^2 + |a_R|^2)\left(8\ln \frac{m_\mu}{m_e} - 11 \right) 
 + |b_L|^2 + |b_R|^2 -2(a_R^{}b_L^*  + a_L^{} b_R^*  + \text{c.c.}) \Bigg], \label{eq:lfv2}
\end{align}
where the detailed formulae for $b_L^{}$ and $b_R^{}$ are presented in Appendix~\ref{sec:app-lfv}. 
We note that the $\mu \to 3e$ data typically do not further constrain the parameter region allowed by the $\ell \to \ell' \gamma$ data. 

%% Fig.~\ref{fig3} shows one-loop box diagrams contributing to the $\ell \to  \ell'\ell''\ell'''$ type of the LFV decay process. 
%% Among the six possible processes, the branching ratio of $\mu \to 3e$ is most strongly constrained by the data. 
%% In our model, the branching ratio of $\mu \to 3e$ is calculated as 
%% \begin{align}
%% {\cal B}(\mu \to ee\bar{e}) &= \frac{1}{4G_F^2}\left(\frac{1}{16\pi^2} \right)^2 \notag\\
%% & \left[
%% |a_{LRLR}^{} |^2 + | a_{LRRL}^{}|^2 +|a_{LLLL}^{}|^2 +|a_{RRRR}^{}|^2 +|a_{LLRR}^{}|^2 +|a_{RRLL}^{}|^2  \right], 
%% \end{align}
%% where $a_{ijkl}^{}$ ($i,j,k,l = L$ or $R$) denote contributions from the diagrams with 
%% the $i$-handed muon and the $j$-handed electron in the $\mu$--$e$ current, 
%% and the $k$-handed positron and the $l$-handed electron in the $e$--$e$ current. 
%% Detailed expressions for each contribution are given in Appendix~\ref{sec:app-lfv}. 

Now, let us numerically show the prediction of the branching ratios of $\ell \to \ell'\gamma$ and $\mu \to 3e$ decays in the parameter sets which satisfy the 
current neutrino data. 
As aforementioned in Sec.~\ref{sec:neutrino}, we take $C_e = 0$ (or $\theta_H = 0$) in the numerical evaluation. 
We assume the CP-conservation in the Yukawa interaction terms. 
The $W$ matrix given in Eq.~(\ref{eq:int2}) becomes the orthogonal matrix which can be parameterized 
by the three angles entered in the $W$ matrix, 
\begin{align}
W = \begin{pmatrix}
1 & 0 & 0 \\ 
0 & \cos w_{23} & -\sin w_{23} \\
0 & \sin w_{23} & \cos w_{23}
\end{pmatrix}
\begin{pmatrix}
\cos w_{13} & 0 & -\sin w_{13} \\
0          & 1 & 0 \\
\sin w_{13} & 0 & \cos w_{13}
\end{pmatrix}
\begin{pmatrix}
\cos w_{12} & -\sin w_{12} & 0 \\
\sin w_{12} & \cos w_{12} & 0 \\
0          & 0 & 1 
\end{pmatrix}. 
\end{align}
We then take the following parameters as inputs; 
\begin{align}
&F_{12},~F_{23},~F_{13},~w_{12},~w_{23},~w_{13},~C_E,~M_1,~M_2,~M_3,~m_{\eta^\pm},~m_{\hat{\eta}^\pm},~V, \notag\\
&\tan\beta,~m_{\eta_\ell^{\pm\pm}},~m_{\eta^0},~m_{\hat{H}^\pm}. 
\end{align}
The parameters in the first line are required for the neutrino mass calculation. 
For the calculation of the LFV decays, we also need to specify the parameters in the second line. 

The current upper limits on the branching ratios of LFV decays of charged leptons are given 
at the 90\% confidence level as
\begin{align}
&{\cal B}(\mu\to e\gamma) < 4.2\times10^{-13}~(\text{MEG~\cite{TheMEG:2016wtm}}),\quad 
{\cal B}(\tau\to e\gamma) < 3.3\times10^{-8}~(\text{BaBar~\cite{Aubert:2009ag}}), \notag\\
&{\cal B}(\tau\to\mu\gamma) < 4.4\times10^{-8}~(\text{BaBar~\cite{Aubert:2009ag}}),\quad  
{\cal B}(\mu \to e e e) < 1.0 \times 10^{-12}~~(\text{SHINDRUM~\cite{Bellgardt:1987du}}). 
\end{align}
We refer to the neutrino oscillation data given in Ref.~\cite{deSalas:2017kay}, and we apply the 3$\sigma$ allowed ranges of 
two squared mass differences and three mixing angles to our analysis. 

\begin{figure}[t]
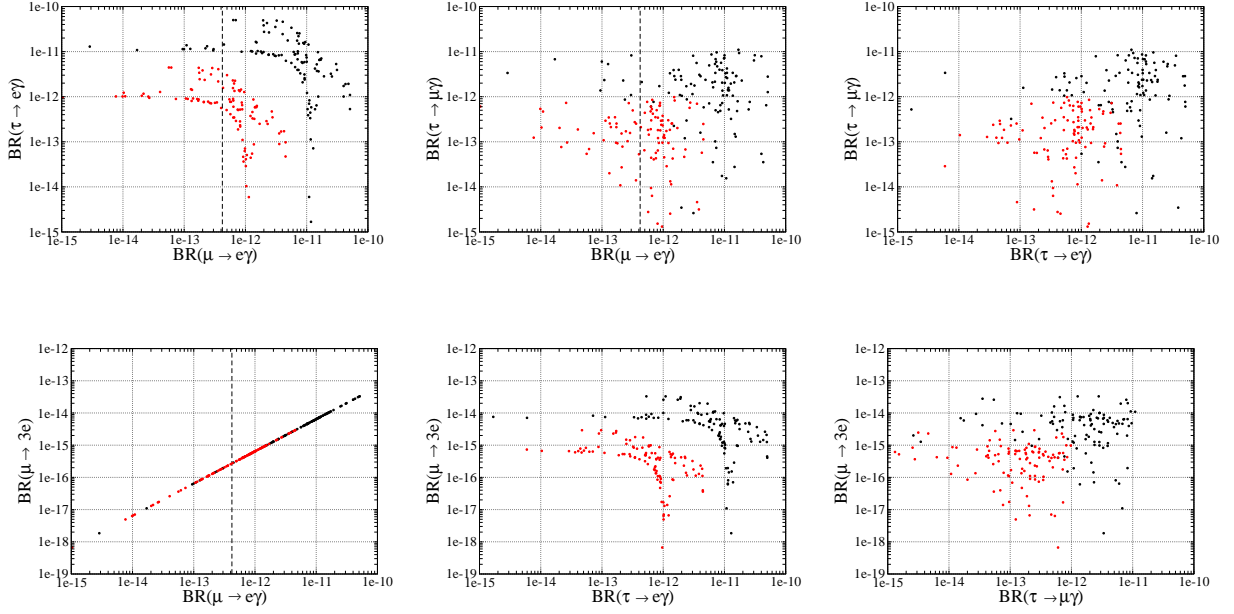

\begin{center}
\includegraphics[width=50mm]{meg-teg.eps} \hspace{3mm} 
\includegraphics[width=50mm]{meg-tmg.eps} \hspace{3mm} 
\includegraphics[width=50mm]{teg-tmg.eps}\\ \vspace{10mm} 
\includegraphics[width=50mm]{meg-m3e-rev.eps}\hspace{3mm} 
\includegraphics[width=50mm]{teg-m3e-rev.eps}\hspace{3mm} 
\includegraphics[width=50mm]{tmg-m3e-rev.eps}
\caption{Correlations between BRs of $\ell \to \ell' \gamma$ (upper figures)
and those of $\ell \to \ell' \gamma$ and $\mu \to 3e$ (lower figures) in the case of $\tan\beta = 30$ (100) for black (red) points. 
The dashed vertical line shows the current upper limit on the branching ratio of $\mu \to e\gamma$. 
In these plots, we fix $V= 10$ TeV, $(M_1,M_2,M_3) = (300,301,302)$ GeV, 
$m_{\eta^\pm}$ = 450 GeV, 
$m_{\hat{\eta}^\pm} = m_{\eta_\ell^{\pm\pm}} = m_{\hat{H}^\pm}$ = 400 GeV and 
$m_{\eta^0}$ = 63 GeV. 
We scan the six parameters of $F_{ij}$ and $w_{ij}$  with the ranges of 
$-10^{-3} \leq F_{ij} \leq 10^{-3}$ and $-\pi/2 \leq w_{ij} \leq \pi/2$. 
All points satisfy the neutrino oscillation data assuming the normal hierarchy case for the neutrino masses.
}
\label{fig6}
\end{center}
\end{figure}

\begin{figure}[t]
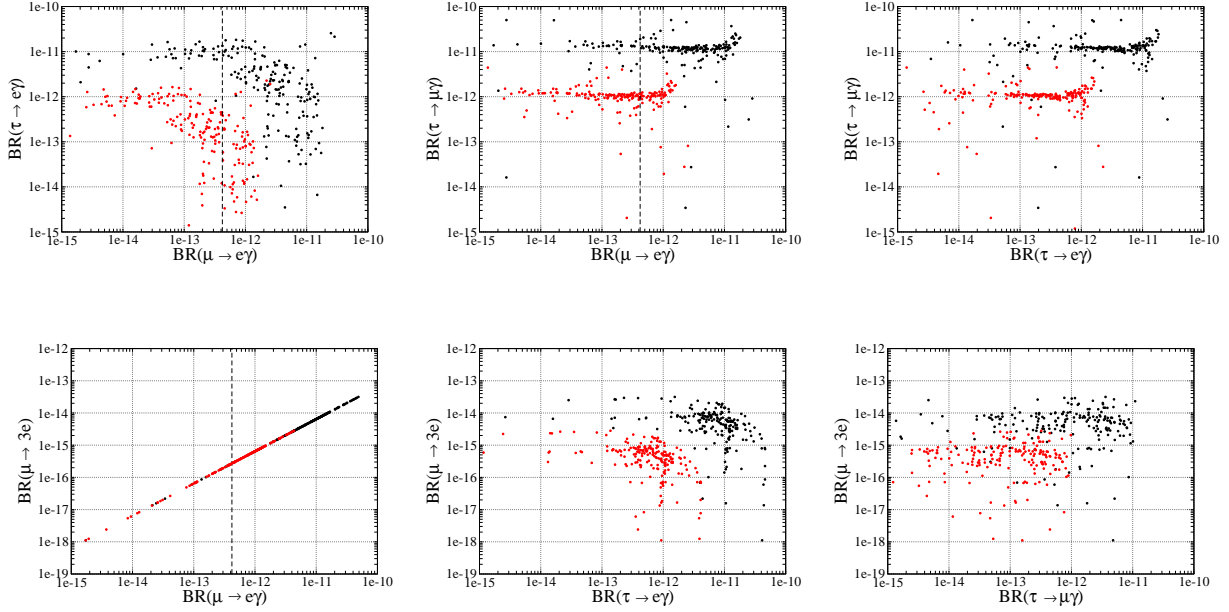

\begin{center}
\includegraphics[width=50mm]{meg-teg-inv.eps} \hspace{3mm} 
\includegraphics[width=50mm]{meg-tmg-inv.eps} \hspace{3mm} 
\includegraphics[width=50mm]{teg-tmg-inv.eps}\\ \vspace{10mm} 
\includegraphics[width=50mm]{meg-m3e-inv-rev.eps}\hspace{3mm} 
\includegraphics[width=50mm]{teg-m3e-inv-rev.eps}\hspace{3mm} 
\includegraphics[width=50mm]{tmg-m3e-inv-rev.eps}
\caption{Same as Fig.~\ref{fig6}, but for the inverted hierarchy case for the neutrino masses. 
}
\label{fig7}
\end{center}
\end{figure}

In Fig.~\ref{fig6}, we show various correlations between the branching ratios of the LFV decays. 
In these plots, we scan six parameters $F_{ij}$ and $w_{ij}$, and fix other parameters as written in the caption of Fig.~\ref{fig6}. 
The coefficient $C_E$ is determined so as to reproduce the mass squared difference of atmospheric neutrinos, i.e., $\Delta m_{\rm atm}^2 = |(m_\nu^3)^2 - (m_\nu^1)^2|$. 
All these points satisfy the current neutrino oscillation data assuming the normal hierarchy for neutrino masses, where the black (red) points show the case with $\tan\beta = 30~(100)$\footnote{We confirm that 
all the relevant parameters in the Lagrangian are enough small in order to ensure the perturbativity.  
In fact, the typical magnitudes of the elements of new Yukawa matrices $y_E^{}$ and $F$ are given by ${\cal O}(10^{-2})$ and ${\cal O}(10^{-3})$, respectively.
In addition, that of the coefficient $C_E$ is found to be 
${\cal O}(10^{-4})$. This can be realized by taking $s_{\theta_\eta}={\cal O}(10^{-2})$, or equivalently taking $\xi_2 ={\cal O}(10^{-2}) $. 
}.
We see that the red points are given in the lower-left region of these planes as compared with the black points, 
because values of the branching ratios of $\ell \to \ell'\gamma$ are dominantly determined by the second term of Eq.~(\ref{ar1}), which is proportional to $c_\beta$. 
We note that the loop contributions of $\eta_\ell^{\pm\pm}$ and $\hat{H}^{\pm}$ are unimportant as long as $|F_{ij}|$ become larger, whose magnitude is typically taken to be smaller than ${\cal O}(10^{-3})$. 
We see that the $\mu \to e\gamma$ data give the most sever constraint on the parameter space, because of its strongest upper bound on the branching ratio. 
For the other two modes $\tau\to \mu\gamma$ and $\tau\to e\gamma$, 
our predictions are typically smaller than the current limit by two or more orders of magnitude, 
because the branching ratio has already been highly suppressed by the $\mu \to e\gamma$ data. 
It is also seen that the branching ratio of $\mu \to 3 e$  is significantly lower than the current upper limit. 
We observe an anticorrelation between the branching ratios of $\mu \to e\gamma$ and $\tau \to e\gamma$, 
see the most upper-left panel, which 
is predicted by our characteristic flavor structure of the Yukawa interactions. 
In addition, we find a very strong correlation between the branching ratios of $\mu \to e\gamma$ and $\mu \to 3e$. 
This can be understood from the fact 
that the $|a_L|^2$ and $|a_R|^2$ terms given in Eq.~(\ref{eq:lfv2}) mainly determine the size of the branching ratio of $\mu \to 3e$.

Similar plots but for the inverted hierarchy case are shown in Fig.~\ref{fig7}. 
We see that ${\cal B}(\tau \to \mu\gamma)$ tends to have similar values with the order of $10^{-11}$ ($10^{-12}$) for $\tan\beta = 30~(100)$ 
as a function of the other branching ratios, which cannot be seen in the normal hierarchy case. 
The other behavior is quite similar to the normal hierarchy case. 

Let us give a comment on cases for the other sets of the fixed parameters in the above analysis. 
Among the fixed parameters, the mass differences between the extra leptons can significantly affect the results of the LFV branching ratios. 
For larger values of the mass difference, these branching ratios tend to become larger, because the suppression by the unitarity of the $W$ matrix becomes weaker. 
Therefore, larger values of $\tan\beta$ or $V$ are required to avoid the constraint from the $\mu \to e\gamma$ data.  
Varying the other mass parameters such as $m_{\eta^\pm}$ does not change the above results significantly. 

\section{Phenomenology \label{sec:pheno} }

In this section, we discuss phenomenological consequences of our model. 
 
\subsection{Dark matter physics \label{sec:dm}}

As we discussed in Sec.~\ref{sec:model}, the lightest $Z_2^{\rm rem}$-odd particle can be a candidate of dark matter; i.e., 
the complex scalar $\eta^0$ or the complex gauge boson $Y^\mu$. 
A scenario with the gauge-boson dark matter $Y^\mu$ is similar to that discussed in Ref.~\cite{Dong:2013wca}, and it has been shown that 
the observed relic abundance is difficult to be explained, due to the too large mass of $Y^\mu$ ($\gtrsim$ 8 TeV). 
We thus consider the scalar boson $\eta^0$ as the dark matter candidate hereafter. 

The scalar boson $\eta^0$ has trilinear interaction terms with neutral $Z_2^{\rm rem}$- and CP-even scalar bosons $H_i$ ($i=1,2,3$), among which $H_1$ can be identified with the discovered Higgs boson ($h$) with the mass of 125 GeV. 
Therefore, the phenomenology of dark matter is similar to the Higgs portal scenario 
in which annihilation processes occur via the $s$-channel Higgs-boson mediations. 

If the dark matter mass $m_{\eta^0}$ is smaller than $2m_W$ and if the additional Higgs bosons $H_2$ and $H_3$ are much heavier than $2m_W$, 
the dominant annihilation process is $\eta^0 \eta^{0*} \to h^{(*)} \to f\bar{f}$ with $f \neq t$, whose thermal averaged cross section is evaluated at the leading order as 
\begin{align}
\langle \sigma v\rangle \simeq \sum_f \frac{m_f}{\pi v}\left(1 - \frac{4m_f^2}{m_{\eta^0}^2} \right)^{3/2}\frac{\lambda_{h\eta\eta}^2}{(4m_{\eta^0}^2 - m_h^2)^2 + m_h^2\Gamma_h^2}, \label{eq:relic}
\end{align}
where $\Gamma_h$ is the width of $h$ ($\sim 4$ MeV) and $\lambda_{h\eta\eta}$ 
is the dimensionful $\eta^0\eta^{0*}h$ coupling.  
From Eq.~(\ref{eq:relic}), the annihilation cross section becomes significant when $m_{\eta^0}$ is getting close to $m_h/2$ due to the resonant effect of $h$, so that 
smaller values of $\lambda_{h\eta\eta}$ are required to keep the observed relic abundance of dark matter, 
$\Omega_{\rm DM}h^2 \simeq 0.12$~\cite{Aghanim:2018eyx}.

\begin{figure}[t]
\begin{center}
\includegraphics[width=80mm]{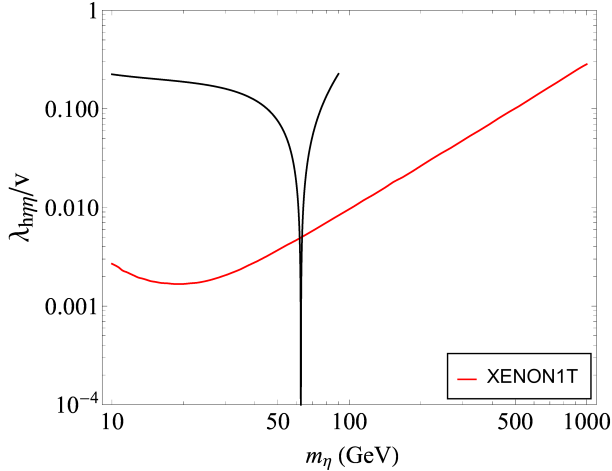}
\caption{Combined results for the dark matter relic abundance and the constraint from the dark matter direct search experiment. 
The red curve shows the upper limit on the normalized coupling by the VEV $\lambda_{h\eta\eta}/v$ as a function of the dark matter mass given by the XENON1T experiment. 
The black curve shows the required value of $\lambda_{h\eta\eta}/v$ satisfying $\Omega_{\rm DM}h^2 = 0.12$ as a function of the dark matter mass. }
\label{fig:dm}
\end{center}
\end{figure}

On the other hand, constraints from dark matter direct detections have to be taken into account. 
In our scenario, the dark matter scattering with a nucleon $N$ through the $t$-channel Higgs mediation becomes 
to be most important. 
Using the effective vertex, which is given by 
\begin{align}
{\cal L}_N = g_N^{}\bar{N}N h, 
\end{align}
with $g_N^{} \simeq 1.1\times 10^{-3}$~\cite{Cheng:2012qr}, the scattering cross section is expressed as
\begin{align}
\sigma_N \simeq \frac{g_N^2 \lambda_{h\eta\eta}^2}{4 \pi (m_N + m_{\eta^0})^2}\frac{m_N^2}{m_h^4}, 
\end{align}
where $m_N$ is the mass of the nucleon. We here neglect the 3-momentum of the dark matter. 

In Fig.~\ref{fig:dm}, we show the combined results for the calculations of the relic abundance and the bound 
from the direct search experiment (XENON1T)~\cite{Aprile:2018dbl}. 
The red curve represents the upper limit on the normalized $h\eta^0\eta^{0*}$ coupling by the VEV $\lambda_{h\eta\eta}/v$ as a function of the dark matter mass $m_{\eta^0}$. 
The required value of $\lambda_{h\eta\eta}/v$ to satisfy  $\Omega_{\rm DM}h^2 = 0.12$ is shown as the black curve. 
As already mentioned, smaller values of $\lambda_{h\eta\eta}$ are required to keep the observed value of the abundance 
when the dark matter mass is around the resonance region $\sim m_h/2$.
Our dark matter candidate can simultaneously satisfy both the relic abundance and the direct detection bounds at $m_{\eta^0} \sim m_h/2$ as it has been known 
in Higgs portal models, see e.g.,~\cite{Kanemura:2010sh,Escudero:2016gzx,Arcadi:2017kky}.  
Another scenario with a much larger mass ($\gtrsim$ a few TeV) may also be considered for the dark matter 
to satisfy both the dark matter data. 
However, we do not discuss details of this case because such a scenario strongly depends on the parameters 
of extra fields. 
Instead, we only have shown that there is at least a solution in our model to satisfy the dark matter data in addition to the neutrino data.

\subsection{Collider physics}

\begin{figure}[t]
\begin{center}
\includegraphics[width=75mm]{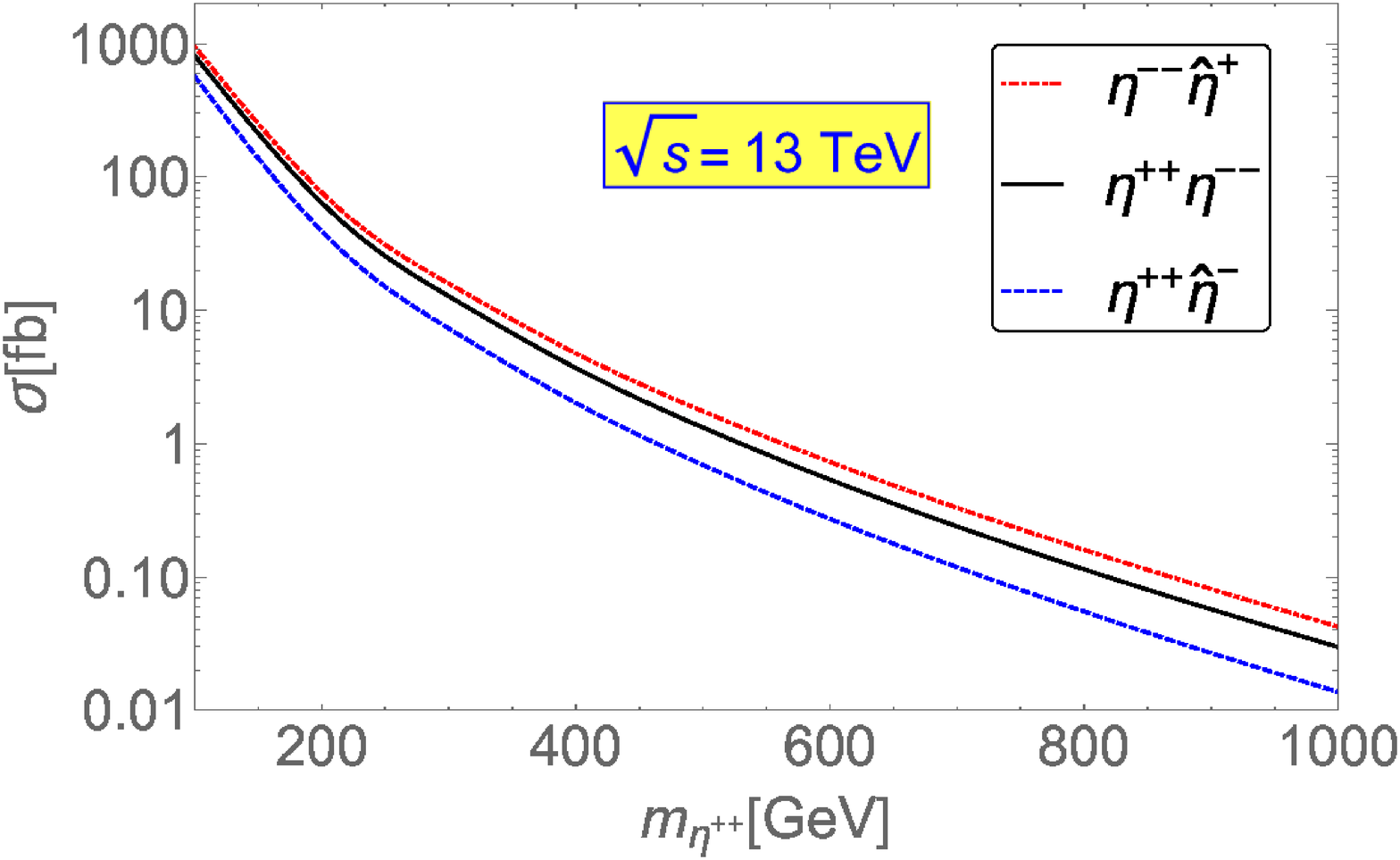}\hspace{3mm}
\includegraphics[width=75mm]{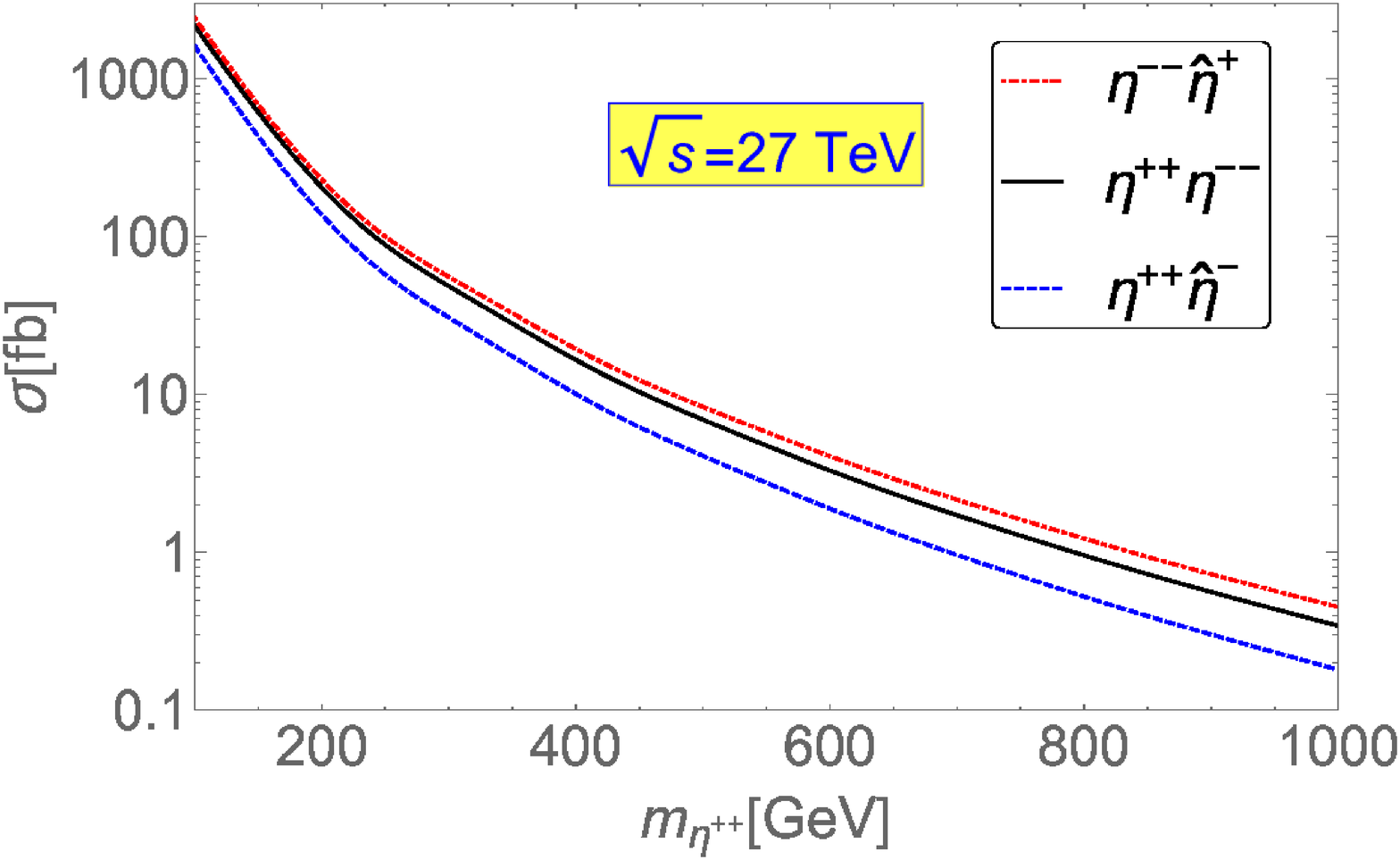}
\caption{Production cross sections for the $pp \to \gamma^*/Z^* \to \eta_\ell^{++}\eta_\ell^{--}$ and $pp \to W^{\pm *} \to \eta_\ell^{\pm\pm}\hat{\eta}^{\mp}$ processes 
as a function of $m_{\eta_{\ell}^{\pm\pm}}$. The collision energy is taken to be 13 TeV (left) and 27 TeV (right).
For the $\eta_\ell^{\pm\pm}\hat{\eta}^{\mp}$ productions, we take $m_{\hat{\eta}^\pm} = m_{\eta_\ell^{\pm\pm} }$. }
\label{fig10}
\end{center}
\end{figure}

\begin{figure}[t]
\begin{center}
\includegraphics[width=75mm]{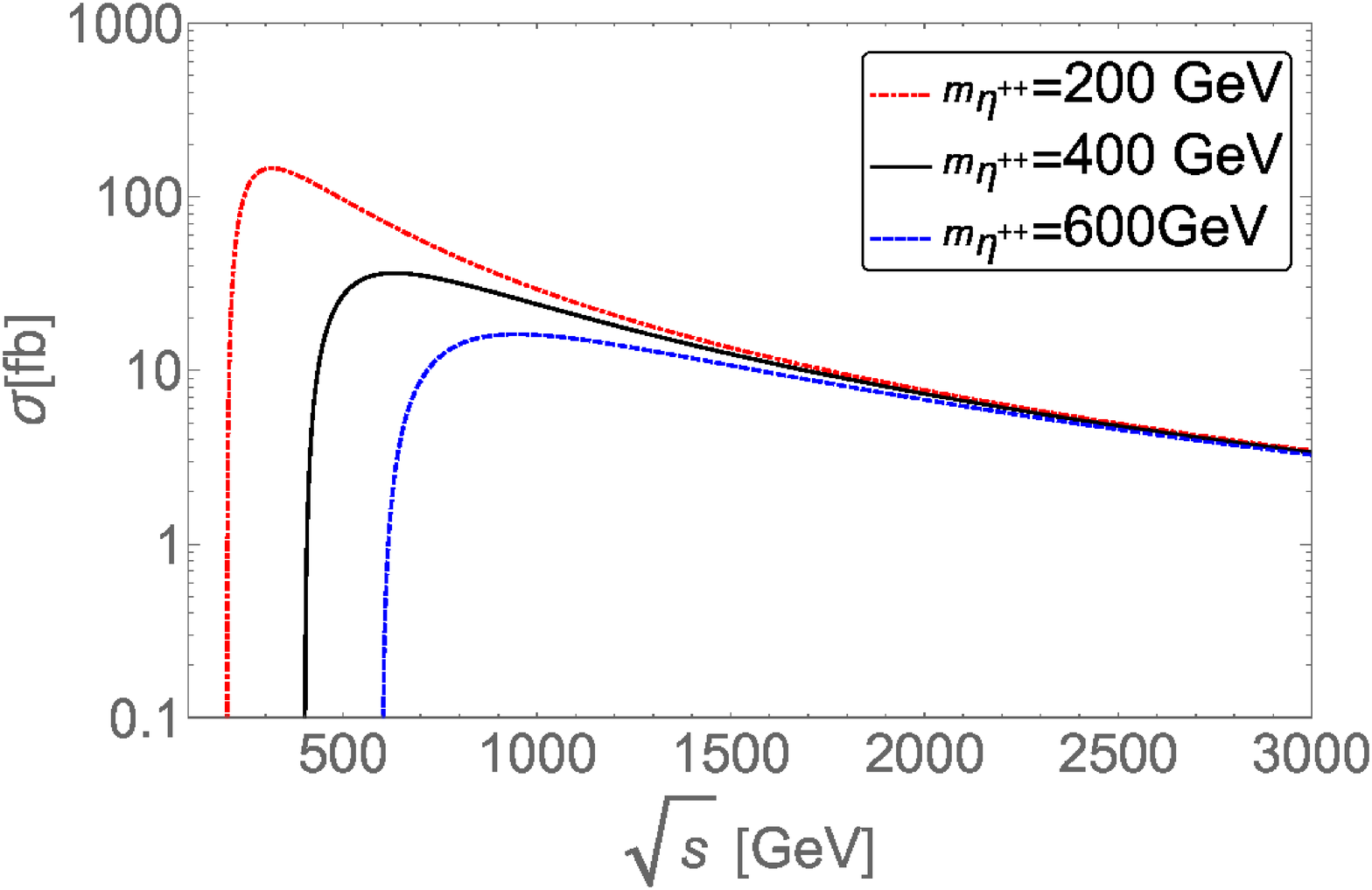}\hspace{3mm}
\includegraphics[width=80mm]{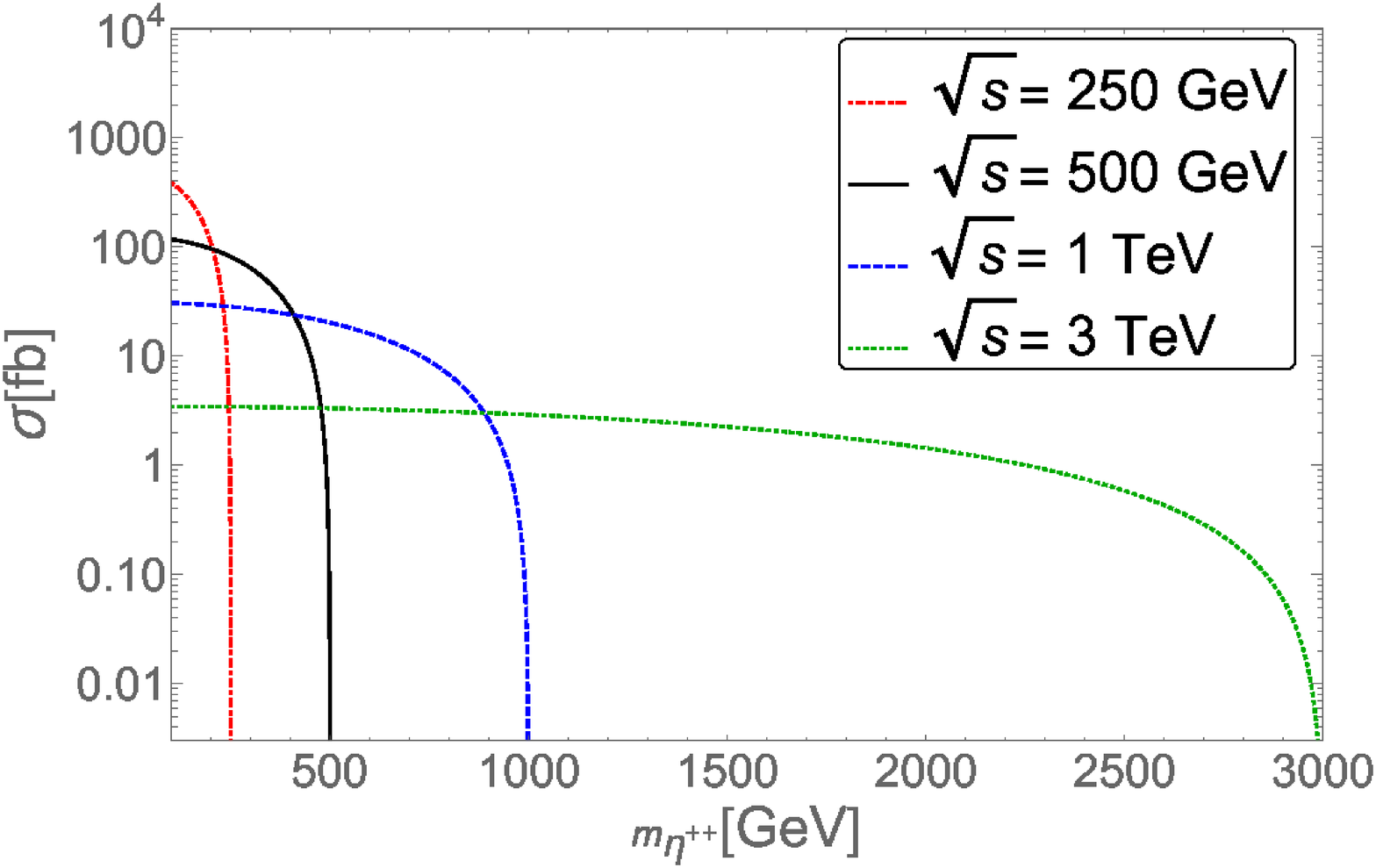}
\caption{Production cross sections for the $e^+e^- \to \gamma^*/Z^* \to \eta_\ell^{++}\eta_\ell^{--}$  process at future lepton colliders. 
The left panel shows the dependence on the center of mass energy $\sqrt{s}$ with fixed values of the mass of $\eta_\ell^{\pm\pm}$ to be 200, 400 and 600 GeV. 
The right panel shows the dependence on the mass of $\eta_\ell^{\pm\pm}$ with fixed values of  $\sqrt{s}$ to be 255 GeV, 500 GeV, 1 TeV and 3 TeV.  }
\label{fig11}
\end{center}
\end{figure}

In our model, there are many new particles, which can potentially be produced at collider experiments. 
However, in the case with $V \gg v$, our model effectively coincides with the THDM with a special Yukawa interaction, which gives extra bosons $H^\pm$, $A$ and $H_2$. 
Because of the special Yukawa interaction, they can decay into quarks with different flavors such as $A/H_2 \to tc$ and $H^\pm \to ts$~\cite{Okada:2016whh}. 
Dedicated simulation studies for these flavor violating decays of the extra Higgs bosons at the LHC have been performed in Ref.~\cite{Gori:2017tvg}.
In addition, there are the other extra particles whose masses are proportional to $V$ such as
$\eta_\ell^{\pm\pm}$, $\eta^{\pm}$, $\hat{\eta}^{\pm}$, $\eta^0$ $\hat{H}^\pm$ and $H_3$. 
They can also be detected at the LHC if the associated coupling constants are small enough.  

One of the most interesting signatures in our model arises from the doubly-charged scalar bosons $\eta_\ell^{\pm\pm}$. 
At the LHC, they can be created in pair via the Drell-Yan process $pp \to \gamma^*/Z^* \to \eta_{\ell}^{++}\eta_{\ell}^{--}$ and in associated with 
the singly-charged scalar bosons $pp \to W^{\pm *} \to \eta_{\ell}^{\pm\pm}\hat{\eta}^{\mp}/\eta_{\ell}^{\pm\pm}\eta^{\mp}$~\cite{Aoki:2011yk}. 
In Fig.~\ref{fig10}, we show the cross sections for these production processes at the $pp$ colliders 
with the collision energy of 13 TeV (left) and 27 TeV (right). 
We use {\tt NNPDF2.3-LO}~\cite{Ball:2012cx} for the parton distribution functions. 
We here neglect effects of the mixing angle $\theta_\eta$ on the associated production cross section, 
because by the analyses for neutrino masses in Sec.~\ref{sec:lfv} 
we have typically $\theta_\eta = {\cal O}(10^{-2})$ which is sufficiently small.
It can be seen that the cross section for the pair production with $m_{\eta_\ell^{\pm\pm}} \simeq 400$ GeV can be a few (a few tens of) fb at $\sqrt{s} =$13 TeV (27 TeV).  
Slightly larger (smaller) values are obtained for the cross section of the associated production $\eta_{\ell}^{++}\hat{\eta}^{-}$ ($\eta_{\ell}^{--}\hat{\eta}^+$). 
In Fig.~\ref{fig11}, we also show the pair production cross section at future lepton colliders, $e^+e^- \to \gamma^*/Z^* \to \eta_{\ell}^{++}\eta_{\ell}^{--}$, as a function of 
the center of mass energy $\sqrt{s}$ (left) and the mass $m_{\eta_\ell^{\pm\pm}}$ (right).

Using the mass spectrum assumed in the analyses for the neutrino masses and LFV decays, 
the decay pattern of $\eta_{\ell}^{\pm\pm}$ is determined to $\eta_{\ell}^{\pm\pm} \to E_k^{\pm}\ell_i^{\pm} \to \ell_i^{\pm}\ell_j^{\pm}\eta^0$, where 
$\eta^0$ is the dark matter candidate.  The intermediate state $E_k^\pm$ can be on-shell in this case. 
The flavor of the same-sign dilepton in the final state is determined by the $F$ and $W$ matrices, which 
are constrained by the neutrino oscillation data and the LFV data. 
%%%
Therefore, future measurements of the flavor of the same-sign dilepton system can give constraints on the structure of the Yukawa interaction in our model, 
and it might be able to provide a hint for the mechanism for the neutrino mass generation. 

The partial decay rates of $E_i^\pm$ and $\eta_{\ell}^{\pm\pm}$ are calculated as 
\begin{align}
\Gamma(E_i^\pm \to \ell_j^\pm \eta^0) &= \frac{M_E}{32\pi}\left(1 - \frac{m_{\eta^0}^2}{M_E^2}\right)^2\left[|(h_R)_{ij}|^2 + |(h_L)_{ij}|^2 \right], \\
\Gamma(\eta_{\ell^{\pm\pm}} \to \ell_i^\pm \ell_j^\pm\eta^0) &= \frac{m_{{\eta_\ell}^{\pm\pm}}}{16\pi}\frac{M_E}{32\pi \Gamma_E}
\left(1 - \frac{m_{\eta_\ell^{\pm\pm}}^2}{M_E^2}\right)^2\left(1 - \frac{m_{\eta^0}^2}{M_E^2}\right)^2\notag\\
& \times \left[|(2W^TFh_R)_{ij}|^2 + |(2W^TFh_L)_{ij}|^2 \right], \label{gam_pp}
\end{align}
where $h_L$ and $h_R$ are given in Eq.~(\ref{flfr}), and the mass of the SM charged leptons is neglected. 
In these expressions, small differences of the masses and the widths of $E_i$~\footnote{The small mass difference is required in order to reproduce the neutrino mixing data and to avoid the constraints from LFV decays of charged leptons.} 
are ignored; i.e., $M_E \equiv M_1 (= M_2 = M_3)$ and $\Gamma_E \equiv \Gamma_{E_1} (= \Gamma_{E_2} = \Gamma_{E_3})$. 
The total width $\Gamma_E$ is typically of order 0.1 GeV, so that the narrow width approximation is valid for the calculation of the decay rate of $\eta_{\ell}^{\pm\pm}$. 

\begin{table}[!t]
\begin{center}
\begin{tabular}{|cccccc||ccccccc|}\hline
     \multicolumn{6}{|c||}{Inputs}  &  \multicolumn{7}{c|}{Outputs} \\\hline\hline
     $F_{12}\times 10^4$   & $F_{23}\times 10^4$     & $F_{13}\times 10^4$   & $w_{12}$  & $w_{23}$  & $w_{13}$ &
     ${\cal B}_{ee}$ & ${\cal B}_{\mu\mu}$ & ${\cal B}_{\tau\tau}$ & ${\cal B}_{e\mu}$ & ${\cal B}_{\mu\tau}$ & ${\cal B}_{e\tau}$ & $\Gamma_{\rm tot}$  \\\hline\hline
     $-$3.03 &  8.13 &  8.44 & $-$1.52 &  $-$0.184 &  0.573 & 12.3 & 1.2 &  5.5 & 15.7 & 34.3  & 31.0 & 8.48 \\\hline
    $-$1.36  &$-$3.99  &$-$4.22 &$-$0.988    &   $-$1.37   &     0.0325 &   0. & 46.9 &  1.6 &  8.7 & 42.4 &  0.4 & 2.08 \\\hline
     0.324 & $-$6.91 &  5.46 & 0.677 & $-$1.39 & $-$0.0841 &   0.1 &   27.8 & 1.7 &  19.1 &  50.3 & 1.0 & 4.78\\\hline
     0.0334 &  5.19 & $-$4.65 &  1.56 & 0.149  &0.958      &    35.8     &     0.2&     7.5&    16.0&     7.7&    32.9&    2.99 \\\hline
     2.11   &  4.35 & 4.09  & $-$1.05 & $-$1.47 & 0.00966 & 0. &46.6&0.5&9.8&42.5&0.5&2.20\\\hline
\end{tabular}
\caption{Benchmark inputs and corresponding outputs, where ${\cal B}_{ij}$ denote ${\cal B}(\eta_{\ell^{\pm\pm}} \to \ell_i^\pm \ell_j^\pm\eta^0)$, 
and $\Gamma_{\rm tot}$ is the total width of $\eta_{\ell^{\pm\pm}}$. 
In this table, $w_{ij}$, ${\cal B}_{ij}$ and $\Gamma_{\rm tot}$ are given in the units of rad, \% and keV, respectively. 
The other input parameters are fixed to be 
$V= 10$ TeV, $(M_1,M_2,M_3) = (300,301,302)$ GeV, 
$m_{\eta^\pm}$ = 450 GeV, 
$m_{\hat{\eta}^\pm} = m_{\eta_\ell^{\pm\pm}} = m_{\hat{H}^\pm}$ = 400 GeV and 
$m_{\eta^0}$ = 63 GeV. 
All these benchmark points satisfy the neutrino oscillation data and the LFV data assuming the normal hierarchy case for the neutrino masses.  }
\label{tab:bp}
\end{center}
\end{table}

\begin{table}[!t]
\begin{center}
\begin{tabular}{|cccccc||ccccccc|}\hline
     \multicolumn{6}{|c||}{Inputs}  &  \multicolumn{7}{c|}{Outputs} \\\hline\hline
     $F_{12}\times 10^4$   & $F_{23}\times 10^4$     & $F_{13}\times 10^4$   & $w_{12}$  & $w_{23}$  & $w_{13}$ &
     ${\cal B}_{ee}$ & ${\cal B}_{\mu\mu}$ & ${\cal B}_{\tau\tau}$ & ${\cal B}_{e\mu}$ & ${\cal B}_{\mu\tau}$ & ${\cal B}_{e\tau}$ & $\Gamma_{\rm tot}$  \\\hline\hline
2.41 & $-$0.716 &  4.34 &  1.43 &  0.957 &  0.0947  &  0. & 65.9 &  0. &   1.1 &  33   &   0.1 &   1.19\\\hline
6.81 & $-$0.982 & $-$6.92 & $-$1.08&  $-$0.427 &    1.08      &   1.4 &   1.0 &  13.3  &  21.0      &   7.8 &  55.4   &   3.02\\\hline
2.61 & $-$0.767 &  6.64 & $-$0.197 &   $-$1.41   &     0.0189 &   0. &   3.0 &   0. &  92.7      &   1.8 &  2.4 &   2.76\\\hline
$-$7.02 & 1.308 & $-$5.32 &  1.36    &   0.733    &  0.205   &   0. &  42.8      &   0.1 &   4.2 &  52.9    &   0.1 &   1.85\\\hline
$-$6.33 & $-$0.792 & 4.32 & $-$0.823  &    $-$0.129  &   1.33      &   7.2 &   0. &   5.2 &   0.1 &   0.1 &  87.3 &   1.19\\\hline
\end{tabular}
\caption{Same as Table~\ref{tab:bp}, but for the inverted hierarchy case. 
 }
\label{tab:bp2}
\end{center}
\end{table}

In Table~\ref{tab:bp} and Table~\ref{tab:bp2}, we give several benchmark points which satisfy the neutrino data and the LFV data for the cases of the normal and the inverted hierarchies, respectively. 
The other input parameters are fixed to be $V= 10$ TeV, $(M_1,M_2,M_3) = (300,301,302)$ GeV, 
$m_{\eta^\pm}$ = 450 GeV and $m_{\hat{\eta}^\pm} = m_{\eta_\ell^{\pm\pm}} = m_{\hat{H}^\pm}$ = 400 GeV.
The dark matter mass $m_{\eta^0}$ is fixed to be 63 GeV in order to satisfy the relic abundance and the constraint from the direct search experiment, see Sec.~\ref{sec:dm}. 
For each point, we show our predictions for the branching ratios of $\eta_{\ell}^{\pm\pm}$ and its total width $\Gamma_{\rm tot}$. 
It is seen that the width of $\eta_{\ell}^{\pm\pm}$ is typically given to be of the order of keV, because the couplings $F_{ij}$ are taken to be ${\cal O}(10^{-3}$--$10^{-4})$ in order to 
avoid the constraint from LFV decays of charged leptons. 
Depending on these benchmark points, $\eta_{\ell}^{\pm\pm}$ can predominantly decay into the same-sign dilepton with various combinations of their flavors. 
Because the decay of $\eta_{\ell}^{\pm\pm}$ contains missing energies which are carried by the dark matter $\eta^0$, 
the invariant mass distribution of the same-sign dilepton system does not have a peak at around the mass of $\eta_{\ell}^{\pm\pm}$. 
This property is different from that of doubly-charged Higgs bosons from an $SU(2)_L$ triplet or a singlet field, which can decay into the same-sign dilepton without missing energy. 
Therefore, the current bounds on the mass of such doubly-charged Higgs bosons, around 800 GeV depending on the flavor of the final state leptons at the LHC~\cite{Aaboud:2017qph}, cannot be applied to 
that on $\eta_{\ell}^{\pm\pm}$. 
In order to extract the bound on the mass of $\eta_{\ell}^{\pm\pm}$ from current experiments and its discovery potential at future experiments, 
dedicated simulation studies are needed, which are beyond the scope of this paper.  

For the sake of completeness, let us discuss the phenomenology for the singly-charged scalar bosons $\eta^\pm$, $\hat{\eta}^\pm$ and $\hat{H}^\pm$. 
For simplicity, we neglect the effect of the small mixing angle $\theta_\eta$ $(\theta_H)$ between $\eta^\pm$ and $\hat{\eta}^\pm$ ($H^\pm$ and $\hat{H}^\pm$). 
These bosons can be produced in pair via the Drell-Yan process at collider experiments. 
In addition, $\hat{\eta}^\pm$ can also be produced in association with $\eta_\ell^{\pm\pm}$ as already discussed in the above text, see also Fig.~\ref{fig10} for its production cross section. 
Their decay processes can be $\eta^\pm/\hat{\eta}^\pm \to \nu_i E_k^\pm  \to \nu_i\ell_j^\pm\eta^0$~\footnote{If $\eta_\ell^{\pm\pm}$ are lighter than $\hat{\eta}^\pm$, the $\hat{\eta}^\pm \to \eta_\ell^{\pm\pm}W^{\mp (*)}$ processes are also possible. } and $\hat{H}^\pm \to \ell_i^\pm\nu_j$. 
Because of the $Z_2^{\rm rem}$ symmetry, the decays of $\eta^\pm$ and $\hat{\eta}^\pm$ include the dark matter. 
The decay of $\eta^\pm$ ($\hat{\eta}^\pm$ and $\hat{H}^\pm$) occurs via the Yukawa coupling $y_E$ ($F$), so that the flavor dependence of the charged lepton in the final state could be different among the decaying particles. 
We note that the singly-charged scalar bosons in the inert doublet model can decay into the W boson and a lighter $Z_2$-odd scalar particle. 
Therefore, the signatures from the decays of $\eta^\pm$ and $\hat{\eta}^\pm$ can be different from those in the inert doublet model~\cite{Aoki:2013lhm,Belyaev:2016lok}. 
On the other hand, the decay property of $\hat{H}^\pm$ is quite similar to that of singly-charged scalar bosons in the Zee model~\cite{Kanemura:2000bq,Cao:2017ffm,Cao:2018ywk}.  

\section{Conclusions \label{sec:conclusion}}

We have proposed a new model for the generation of tiny neutrino masses based on the  
3-3-1 gauge symmetry within the minimal fermion content required for the gauge anomaly cancellation. 
In this model, the source for lepton number violation is obtained by extending the minimal Higgs sector of 3-3-1 models with an additional $SU(3)_L$ triplet scalar field.  
Majorana masses for the active neutrinos are generated at one-loop level. 
We have found the parameter sets which satisfy the current neutrino data under the constraint from the LFV decays 
of the charged leptons such as $\mu \to e\gamma$ and $\mu \to 3e$.  

In our model, $Z_2^{\rm rem}$ appears as a remnant symmetry after the breaking of the electroweak symmetry and that of the global $U(1)'$ symmetry, 
where the latter symmetry is introduced to avoid the dangerous flavor changing neutral current. 
The symmetry $Z_2^{\rm rem}$ guarantees the stability of the dark matter candidate which is the lightest neutral $Z_2^{\rm rem}$-odd scalar particle $\eta^0$. 
We have confirmed that the dark matter candidate can satisfy the relic abundance and the direct search results when the dark matter mass is taken to be at around 
the half of the discovered Higgs boson mass.  

We then have discussed the collider phenomenology of our model. 
One of the most interesting signatures arise from productions and decays of the doubly-charged scalar bosons $\eta_\ell^{\pm\pm}$, because of 
the characteristic flavor dependence of the same-sign dilepton in the final state. 
Even if $\eta_\ell^{\pm\pm}$ are too heavy to be detected at collider experiments, 
the Higgs sector of our model, which effectively coincides with the THDM at $V \gg v$, 
predicts the special structure of the Yukawa interaction to the SM quarks. 
The extra Higgs bosons can then mainly decay into quarks with different flavors, so that 
the detection of such bosons would be important to test our model. 

In conclusion, our model can give an interesting testable example of the 3-3-1 scenario, 
where the number of generation of quarks and leptons, neutrino oscillation and dark matter can be explained simultaneously.

\begin{acknowledgments}

The authors would like to thank Niko Koivunen and Katri Huitu for useful discussions.  
The works of A.~D. and S.~K. were supported in part by JSPS, Grant-in-Aid for Scientific Research, No. 18F18321. 
K.~E. was supported in part by the Sasakawa Scientific Research Grant from The Japan Science Society. 
S.~K. was supported in part by Grant-in-Aid for Scientific Research on Innovative Areas, the Ministry of Education,
Culture, Sports, Science and Technology, No. 16H06492 and No. 18H04587, and also by
JSPS, Grant-in-Aid for Scientific Research, Grant No. 18F18022.
The work of K.~Y. was supported in part by the Grant-in-Aid for Early-Career Scientists, No.~19K14714. 

\end{acknowledgments}

\newpage

\begin{appendix}

\section{Stationary conditions\label{sec:stationary}}

We show that the vanishment of the VEV $v'$ given in Eq.~(\ref{eq:vev}) is guaranteed due to the remnant $\tilde{Z}_2$ symmetry which is discussed in Sec.~\ref{sec:model}. 
The stationary conditions for each neutral component of the scalar field can be expressed as
\begin{align}
\frac{\partial V}{\partial \varphi^0}\Big|_0 = 0, 
\end{align}
where $|_0$ represents taking all the fields to be zero after the derivative. 
We note that nonzero values of the left-hand side appear for the real component of the scalar field due to the assumption of CP-conservation in the Higgs potential. 
For the condition of the real component of $\eta_1^0$, we obtain 
\begin{align}
m_{13}^2V + \frac{v'v}{2}(\sqrt{2}\mu s_\beta + V\rho_{13} c_\beta) = 0. 
\end{align}
Because the $\tilde{Z}_2$ symmetry in the Lagrangian forbids the $m_{13}^2$ term, the above equation is satisfied by taking $v' = 0$ with $v\neq 0$ for arbitrary nonzero values of the term inside the parenthesis. 
In this case, the condition for the real component of $\eta_3^0$ is simultaneously satisfied, and the non-trivial conditions appear from those for the real components of $\phi_{a}^0$ ($a = 1,2,3$) as follows: 
\begin{align}
& v \left[2 m_1^2c_\beta +v^2c_\beta(2\lambda_1 c^2_\beta +  \lambda_{12} s^2_\beta ) + V^2 \lambda_{13}c_\beta  - \sqrt{2} V \mu  s_\beta \right] = 0, \\
& v \left[2 m_2^2 s_\beta +v^2 s_\beta \left(2\lambda_2 s^2_\beta + \lambda_{12} c^2_\beta   \right) + V^2 \lambda_{23} s_\beta - \sqrt{2} V \mu  c_\beta \right]= 0, \\
& V \left[2m_3^2 + 2V^2 \lambda_3  + v^2(\lambda_{13} c^2_\beta + \lambda_{23} s^2_\beta)\right] - \sqrt{2} \mu  v^2s_\beta c_\beta  =  0. 
\end{align}
We can solve these equations in terms of $m_1^2$, $m_2^2$ and $m_3^2$.

\section{Mass formulae for the scalar bosons\label{sec:mass}}

We first give the mass formulae for the $Z_2^{\rm rem}$-even scalar bosons. 
The neutral components of the scalar triplet fields can be expressed as 
\begin{align}
\phi_a^0 = \frac{1}{\sqrt{2}}(\phi_a^R + v_a + i \phi_a^I ), \quad (a = 1,2,3), 
\end{align}
with $v_3 = V$. 
There are 3 pairs of singly-charged, 
3 CP-odd and 3 CP-even scalar states in the $Z_2^{\rm rem}$-even sector.
Their mass eigenstates are obtained by introducing the following orthogonal transformations: 
\begin{align}
\begin{pmatrix}
\phi_1^\pm \\
\phi_2^\pm \\
\phi_\ell^\pm 
\end{pmatrix}
&= 
\begin{pmatrix}
c_\beta & s_\beta & 0\\
-s_\beta & c_\beta& 0\\
0&0&1
\end{pmatrix}
\begin{pmatrix}
G^\pm \\
\tilde{\phi}^\pm \\
\hat{H}^{\pm} 
\end{pmatrix}
=
\begin{pmatrix}
c_\beta & s_\beta & 0\\
-s_\beta & c_\beta& 0\\
0&0&1
\end{pmatrix}
\begin{pmatrix}
1&0&0 \\
0&c_{\theta_H} & -s_{\theta_H} \\
0&s_{\theta_H} & c_{\theta_H} \\
\end{pmatrix}
\begin{pmatrix}
G^\pm \\
H^\pm \\
\hat{H}^{\pm} 
\end{pmatrix}, \label{mixing1} \\
\begin{pmatrix}
\phi_1^I \\
\phi_2^I \\
\phi_3^I
\end{pmatrix}
&= ({\bm x}_{12}^+,{\bm x}_{12}^-,{\bm x}_3)
\begin{pmatrix}
G^0 \\
G^{\prime 0} \\
A
\end{pmatrix},  \label{mat-odd} \\
\begin{pmatrix}
\phi_1^R \\
\phi_2^R \\
\phi_3^R
\end{pmatrix}
&= 
R_H
\begin{pmatrix}
H_1 \\
H_2 \\
H_3
\end{pmatrix}, \label{mat-even}
\end{align}
where $G^\pm$, $G^0$ and $G^{\prime 0}$ are the NG bosons which are absorbed into the longitudinal component of $W^\pm$, $Z$ and $Z'$, respectively. 
In Eq.~(\ref{mat-even}), $R_H$ is the $3\times 3$ orthogonal matrix which can be expressed by three independent mixing angles. These mixing angles are determined 
from the mass matrix for the CP-even Higgs bosons given in Eq.~(\ref{eq:cp-even}).  
In Eq.~(\ref{mat-odd}), ${\bm x}_{12}^\pm \equiv ({\bm x}_1 \pm {\bm x}_2)/|{\bm x}_{12}^\pm|$ and ${\bm x}_3$ are three component vectors defined as 
\begin{align}
{\bm x}_1^T &= \left(-\frac{v}{V}c_\beta(1 + \frac{v^2}{V^2}c_\beta^2 )^{-1/2}, 0, (1 + \frac{v^2}{V^2}c_\beta^2 )^{-1/2} \right), \\
{\bm x}_2^T &= \left(c_\beta,-s_\beta,0 \right), \\
{\bm x}_3^T &= \left(s_\beta(1 + \frac{v^2}{V^2}s_\beta^2 c_\beta^2)^{-1/2},  c_\beta (1 + \frac{v^2}{V^2}s_\beta^2c_\beta^2)^{-1/2},(1 + \frac{V^2}{v^2s_\beta^2c_\beta^2} )^{-1/2}\right). 
\end{align}
The squared masses of physical Higgs bosons and the mixing angle $\theta_H$ are given by 
\begin{align}
m_{H^\pm}^2 &= c_{\theta_H}^2({\cal M}_{H^\pm}^2)_{11} + s_{\theta_H}^2({\cal M}_{H^\pm}^2)_{22} + s_{2\theta_H}({\cal M}_{H^\pm}^2)_{12}, \\
m_{\hat{H}^\pm}^2 &= s_{\theta_H}^2({\cal M}_{H^\pm}^2)_{11} + c_{\theta_H}^2({\cal M}_{H^\pm}^2)_{22} - s_{2\theta_H}({\cal M}_{H^\pm}^2)_{12}, \\
m_{A}^2 & = \frac{\sqrt{2}\mu V}{s_{2\beta} } \left(1 + \frac{v^2}{V^2}s_\beta^2 c_\beta^2 \right), \\
m_{H_a}^2 & =(R_H^T {\cal M}^2_H R_H)_{aa}, \quad  (a = 1,2,3), \\
\tan2\theta_H &= \frac{2({\cal M}^2_{H^\pm})_{12}}{({\cal M}^2_{H^\pm})_{11}  - ({\cal M}^2_{H^\pm})_{22}}. 
\end{align}
We can identify $m_{H_1}$ as the mass of the discovered Higgs boson, 125 GeV. 
The mass matrices ${\cal M}_{H^\pm}^2$ and ${\cal M}_{H}^2$ are calculated as 
\begin{align}
{\cal M}_{H^\pm}^2 &= \begin{pmatrix}
 \frac{v^2 \rho_{12}}{2} + \frac{ \sqrt{2} V \mu }{s_{2\beta}} & \frac{vV \xi_1 }{2} \\
   & m_4^2+ \frac{V^2 ( \lambda_{34} +  \rho_{34} ) + v^2 (c_\beta^2\lambda_{14} + s_\beta^2  \lambda_{24} )}{2}
\end{pmatrix}, \\
{\cal M}_{H}^2 &= 
\begin{pmatrix}
2v^2\lambda_1c_\beta^2  +\frac{\mu V}{\sqrt{2}}\tan\beta & v^2\lambda_{12}s_\beta c_\beta  - \frac{\mu V}{\sqrt{2}} & vV\lambda_{13} c_\beta  - \frac{\mu V}{\sqrt{2}}s_\beta  \\
 & 2v^2\lambda_2 s_\beta^2 + \frac{\mu V}{\sqrt{2}}\cot\beta & vV\lambda_{23}s_\beta - \frac{\mu v}{\sqrt{2}}c_\beta\\
&& 2V^2\lambda_3 + \frac{\mu v^2}{\sqrt{2}V}c_\beta s_\beta
\end{pmatrix}. \label{eq:cp-even}
\end{align}
In the above expressions, the lower-left elements are the same as the corresponding transposed elements. 

Next, we present the mass formulae for the $Z_2^{\rm rem}$-odd scalar states, in which 
there are one pair of doubly-charged, 3 pairs of singly-charged and one neutral complex scalar states. 
The squared mass of the doubly-charged scalar bosons $\eta_\ell^{\pm\pm}$ is given by 
\begin{align}
m_{\eta_\ell^{\pm\pm}}^2 = m_4^2 + \frac{1}{2} \left[ V^2 \lambda_{34} + v^2(c_\beta^2 \lambda_{14} + s_\beta^2 \lambda_{24} + s_\beta^2\rho_{24}) \right]. 
\end{align}
The mass eigenstates of the singly-charged and neutral states are defined as follows: 
\begin{align}
\begin{pmatrix}
\eta_2^\pm \\
\eta_3^\pm \\
\eta_\ell^\pm 
\end{pmatrix}
&= 
\frac{1}{\sqrt{1 + \frac{v^2s_\beta^2}{V^2}}}
 \begin{pmatrix}
 \frac{v}{V}s_\beta & 1 & 0\\
 -1 & \frac{v}{V}s_\beta & 0\\
 0&0&1
 \end{pmatrix}
 \begin{pmatrix}
 G^{\prime \pm} \\
 \tilde{\eta}^\pm \\
\eta_\ell^\pm 
 \end{pmatrix} \notag\\
&=
\frac{1}{\sqrt{1 + \frac{v^2s_\beta^2}{V^2}}}
\begin{pmatrix}
\frac{v}{V}s_\beta & 1 & 0\\
-1 & \frac{v}{V}s_\beta & 0\\
0&0&1
\end{pmatrix}
\begin{pmatrix}
1&0&0 \\
0&c_{\theta_\eta} & -s_{\theta_\eta} \\
0&s_{\theta_\eta} & c_{\theta_\eta} \\
\end{pmatrix}
\begin{pmatrix}
G^{\prime \pm} \\
\eta^\pm \\
\hat{\eta}^{\pm} 
\end{pmatrix}, \label{mixing2}\\
\begin{pmatrix}
\eta_1^0 \\
\eta_3^{0*} 
\end{pmatrix}
&= 
\frac{1}{\sqrt{1 + \frac{v^2c_\beta^2}{V^2}}}
\begin{pmatrix}
\frac{v}{V}c_\beta  &  1 \\
-1 & \frac{v}{V}c_\beta 
\end{pmatrix}
\begin{pmatrix}
G_Y^0 \\
\eta^{0*}
\end{pmatrix},  \label{eq:dm}
\end{align}
where $G^{\prime \pm}$ and  $G_Y^{0}$ are the NG bosons which are absorbed into the longitudinal component of $W'$ and $Y$, respectively. 
The squared masses of the physical scalar bosons and the mixing angle $\theta_\eta$ are given by 
\begin{align}
m_{\eta^\pm}^2 &= c_{\theta_\eta}^2({\cal M}_{\eta^\pm}^2)_{11} + s_{\theta_\eta}^2({\cal M}_{\eta^\pm}^2)_{22} + s_{2\theta_\eta}({\cal M}_{\eta^\pm}^2)_{12}, \\
m_{\hat{\eta}^\pm}^2 &= s_{\theta_\eta}^2({\cal M}_{\eta^\pm}^2)_{11} + c_{\theta_\eta}^2({\cal M}_{\eta^\pm}^2)_{22} - s_{2\theta_\eta}({\cal M}_{\eta^\pm}^2)_{12}, \\
m_{\eta^0}^2 & = \frac{V^2}{2} \left(1 + \frac{v^2}{V^2}c_\beta^2 \right) \left(\rho_{13} + \sqrt{2} \tan\beta \frac{\mu}{V} \right), \\
\sin2\theta_\eta &= \frac{2({\cal M}^2_{\eta^\pm})_{12}}{m_{\eta^\pm}^2 - m_{\hat{\eta}^\pm}^2}. \label{sin2theta}
\end{align}
The mass matrix for the singly-charged state ${\cal M}_{\eta^\pm}^2$ is calculated as 
\begin{align}
{\cal M}_{\eta^\pm}^2 &= \begin{pmatrix}
 \frac{V^2 }{2}\left(\rho_{23} + \frac{\sqrt{2}\mu}{V\tan\beta} \right)\left(1 + \frac{v^2}{V^2} s_\beta^2\right) & \frac{vV\xi_2 c_\beta}{2}\sqrt{1+ \frac{v^2}{V^2} s_\beta^2}  \\
 \frac{vV\xi_2 c_\beta}{2}\sqrt{1+ \frac{v^2}{V^2} s_\beta^2}  &  m_4^2 + \frac{V^2  \lambda_{34}+ v^2c_\beta^2 (\lambda_{14} + \rho_{14}) + v^2s_\beta^2   \lambda_{24}}{2} 
\end{pmatrix}. 
\end{align}

\section{Formulae for lepton flavor violating decays of charged leptons \label{sec:app-lfv}}

We give the expressions for $a_{L,R}$ and $b_{L,R}$ appearing in the branching ratios of $\ell_i \to \ell_j \gamma$ and $\mu \to 3e$ given in Eqs.~(\ref{eq:lfv1}) and (\ref{eq:lfv2}), respectively. 
They can be separately expressed as
\begin{align}
a_{L,R} = \sum_{\varphi}a_{L,R}^\varphi,\quad 
b_{L,R} = \sum_{\varphi}b_{L,R}^\varphi, 
\end{align}
where $a_{L,R}^\varphi$ and $b_{L,R}^\varphi$ denote the $\varphi$--loop ($\varphi = \eta^0$, $\eta^{\pm\pm}$ and $\hat{H}^\pm$) contribution to the amplitude for the LFV decays. 
Each of them is calculated as follows: 
\begin{align}
(a_R^{\eta^0})_{ij} &=(a_L^{\eta^0})_{ij} =
\frac{1}{16\pi^2} \sum_k \Bigg[
 \frac{(h_L^*)_{kj} (h_L^{})_{ki}}{2M_k^2}G_2\left(\frac{m_{\eta^0}^2}{M_k^2}\right)
 + \frac{(h_L^*)_{kj} (h_R^{})_{ki}}{M_km_{\ell_i}} G_1\left(\frac{m_{\eta^0}^2}{M_k^2}\right)  \Bigg], \label{ar1}  \\
(a_R^{\eta^{\pm\pm}})_{ij} &=
-\frac{1}{4\pi^2} \sum_k (W^TF^*)_{kj} (W^TF^{})_{ki} \left[
 \frac{1}{2M_k^2}G_2\left(\frac{m_{\eta^{\pm\pm}}^2}{M_k^2}\right) + \frac{1}{m_{\eta^{\pm\pm}}^2}G_2\left(\frac{M_k^2}{m_{\eta^{\pm\pm}}^2}\right) \right], \label{ar2} \\
(a_R^{\hat{H}^\pm})_{ij} &=
-\frac{1}{4\pi^2}\frac{1}{12m_{\hat{H}^{\pm}}^2} \sum_k F^*_{kj} F_{ki},  \\
(a_L^{\eta^{\pm\pm}})_{ij} &= (a_L^{\hat{H}^\pm})_{ij} = 0, \label{ar4} 
\end{align}
\begin{align}
(b_{L,R}^{\eta^0})_{ij} & =
-\frac{1}{12\pi^2} \sum_k \frac{(h_{L,R}^*)_{kj} (h_{L,R}^{})_{ki}}{8M_k^2}G_3\left(\frac{m_{\eta^{0}}^2}{M_k^2} \right), \\
(b_L^{\eta^{\pm\pm}})_{ij} & =
\frac{1}{12\pi^2} \sum_k(W^TF)_{kj}^* (W^TF^{})_{ki}\Bigg[\frac{1}{2M_k^2}G_3\left(\frac{m_{\eta^{\pm\pm}}^2}{M_k^2} \right)
+\frac{1}{m_{\eta^{\pm\pm}}^2}G_4\left(\frac{M_k^2}{m_{\eta^{\pm\pm}}^2} \right)\Bigg], \\
(b_L^{\hat{H}^\pm})_{ij} & =
\frac{1}{12\pi^2}\frac{1}{6m_{\hat{H}^{\pm}}^2} \sum_k F^*_{kj} F_{ki}, \\
(b_R^{\eta^{\pm\pm}})_{ij} & = (b_R^{\hat{H}^\pm})_{ij} = 0, 
\end{align}
where $h_L$ and $h_R$ are the coefficients of the $\bar{E}_R'\ell_L' \eta^{0*}$ and $\bar{E}_L'\ell_R' \eta^{0*}$ vertex, respectively, given as 
\begin{align}
h_L = \frac{\sqrt{2}}{\sqrt{1 + v^2 c_\beta^2/V^2}}\frac{v}{V^2}c_\beta M_E^{\rm diag}W^\dagger,\quad 
h_R = \frac{\sqrt{2}}{\sqrt{1 + v^2 c_\beta^2/V^2}}\frac{1}{v} W^\dagger M_e^{\rm diag}. \label{flfr}
\end{align}
The loop functions $G_{1,2,3,4}$ are given by 
\begin{align}
G_1(x) &=\frac{1-4x+3x^2-2x^2\ln x}{2(1-x)^3},\\
G_2(x) &=\frac{1-6x+3x^2+2x^3-6x^2\ln x}{6(1-x)^4}, \\
G_3(x) &=\frac{-7+36x-45x^2+16x^3+18x^2\ln x-12x^3\ln x}{6(1-x)^4}, \\
G_4(x) &=\frac{2-9x+18x^2-11x^3+6x^3\ln x}{6(1-x)^4}. 
\end{align}

\end{appendix}

\vspace*{4mm}

%\BibitemShut{NoStop}
\bibliography{references}

\end{document}